\documentclass[12pt,amssymb]{article}
\usepackage[usenames,dvipsnames]{color}

\usepackage{graphicx}
\usepackage{epstopdf}
\usepackage{amsfonts,amsmath}

\newcommand{\bea}{\begin{eqnarray}}

\newcommand{\eea}{\end{eqnarray}}

\newcommand{\beq}{\begin{equation}}

\newcommand{\eeq}{\end{equation}}

\newcommand{\lav}{\langle}

\newcommand{\rav}{\rangle}

\begin{document}

\def \bea{\begin{eqnarray}}
\def \eea{\end{eqnarray}}
\def \ua{{\uparrow}}
\def \da{{\downarrow}}
\def \tr{{\mbox{tr~}}}
\def \ra{{\rightarrow}}
\def \ua{{\uparrow}}
\def \da{{\downarrow}}
\def \be{\begin{equation}}
\def \ee{\end{equation}}
\def \ba{\begin{array}}
\def \ea{\end{array}}
\def \bea{\begin{eqnarray}}
\def \eea{\end{eqnarray}}
\def \nn{\nonumber}
\def \l{\left}
\def \r{\right}
\def \half{{1\over 2}}
\def \etal{{\it {et al}}}
\def \cH{{\cal{H}}}
\def \cM{{\cal{M}}}
\def \cN{{\cal{N}}}
\def \cQ{{\cal Q}}
\def \cI{{\cal I}}
\def \cV{{\cal V}}
\def \cG{{\cal G}}
\def \cF{{\cal F}}
\def \cZ{{\cal Z}}
\def \bS{{\bf S}}
\def \bI{{\bf I}}
\def \bL{{\bf L}}
\def \bG{{\bf G}}
\def \bQ{{\bf Q}}
\def \bK{{\bf K}}
\def \bR{{\bf R}}
\def \br{{\bf r}}
\def \bu{{\bf u}}
\def \bq{{\bf q}}
\def \bk{{\bf k}}
\def \bz{{\bf z}}
\def \bx{{\bf x}}
\def \bpsi{{\bar{\psi}}}
\def \tJ{{\tilde{J}}}
\def \W{{\Omega}}
\def \e{{\epsilon}}
\def \lam{{\lambda}}
\def \L{{\Lambda}}
\def \a{{\alpha}}
\def \t{{\theta}}
\def \b{{\beta}}
\def \g{{\gamma}}
\def \D{{\Delta}}
\def \d{{\delta}}
\def \w{{\omega}}
\def \s{{\sigma}}
\def \f{{\varphi}}
\def \x{{\chi}}
\def \e{{\epsilon}}
\def \h{{\eta}}
\def \G{{\Gamma}}
\def \z{{\zeta}}
\def \hatt{{\hat{\t}}}
\def \hn{{\bar{n}}}
\def \vk{{\bf{k}}}
\def \vq{{\bf{q}}}
\def \gk{{\g_{\vk}}}
\def \nd{{^{\vphantom{\dagger}}}}
\def \yd{^\dagger}
\def \av#1{{\langle#1\rangle}}
\def \ket#1{{\,|\,#1\,\rangle\,}}
\def \bra#1{{\,\langle\,#1\,|\,}}
\def \braket#1#2{{\,\langle\,#1\,|\,#2\,\rangle\,}}

\Large
\noindent \textcolor{NavyBlue}{\bf Ultracold Atomic Gases: Novel States of Matter}

\setcounter{secnumdepth}{0}
\normalsize
\bigskip
\bigskip
\noindent LUDWIG MATHEY$^a$, SHAN-WEN TSAI$^b$, and ANTONIO H. CASTRO NETO$^c$

\bigskip

\noindent $^a$ Harvard University, Cambridge, Massachusetts, USA\\
$^b$ University of California, Riverside, California, USA\\
$^c$ Boston University, Boston, Massachusetts, USA\\

\bigskip
\medskip

\tableofcontents


\bigskip

\section*{\textcolor{NavyBlue}{\bf Glossary}}
\addcontentsline{toc}{section}{\bf Glossary}
\normalsize


\noindent {\bf Bose-Einstein Condensation (BEC)}\\
Low temperature phase of systems of identical bosons,  characterized
 by superfluidity.\\

\noindent{\bf Boson}\\
Particles with integer spin $S=0,1,2,...$. Mediators of interactions, such as photons and gluons are bosons. Objects made of an even number of fermions are bosons: positroniun (electron + positron), meson (two quarks), $^{87}$Rb (37 protons, 48 neutrons and 37 electrons), $^7$Li (3 protons, 4 neutrons, 3 electrons).\\


\noindent {\bf Cooper pairs}\\
 At low temperatures and for attractive interactions
 fermions form a superconducting state, in which
 fermions form pairs which condense.\\

\noindent {\bf Fermi surface}\\
Since fermions obey Pauli's exclusion principle, the ground state of $N$ non-interacting fermions in $d$-dimensions is the state with the $N$ lowest energy states occupied. In momentum space the last occupied state and the first unoccupied state define a surface of dimensions $d-1$, called the Fermi surface. \\

\noindent {\bf Fermion}\\
Particles with half-odd integer spin $S=\frac1 2, \frac 3 2, \frac 5 2,...$. Examples include elementary particles such as electrons and quarks. Objects made of an odd number of fermions are also fermionic, such as protons, $^{40}$K (19 protons, 21 neutrons, and 19 electrons), and $^6$Li (3 protons, 3 neutrons, and 3 electrons).\\

\noindent{\bf Laser Cooling}\\
 In a typical experimental setup, the atoms
 are cooled to the regime of $10^2 \mu K$, by using
 pairs of counterpropagating laser beams 
 that are slightly red-detuned below an atomic transition.
 Due to the Doppler effect the atoms can only absorb a photon if
 they travel towards the beam with a
 high velocity. From that process the atoms experience
 a recoil, which slows them down.
\\ 

\noindent{\bf Evaporative Cooling} \\
 To slow the atoms down further, to the $\mu K$ regime,
 one applies radio frequency radiation
 that flips the internal state
 to a high-field seeking, i.e. non-trapped, state in such a way,
 that only atoms of high kinetic energy
 can escape. Due to thermalization, this leads
 to cooling of the remaining atomic ensemble.
\\

\noindent {\bf Magnetic Trap}\\
 The atoms are trapped by applying 
 a spatially inhomogeneous magnetic field. This
 field leads to an energy shift due to the Zeeman effect,
 which the atoms experience as an external potential, for
 large energy splittings of the magnetic levels.
 Different geometric designs are in use, such as the TOP trap,
 or the Ioffe-Pritchard trap.\\

\noindent {\bf Optical Lattice}\\
 Counterpropagating laser beams create a standing wave field, which
 the atoms experience as a periodic potential, due to
 the ac Stark shift.
 If the temperature and all energy scales are small compared
 to the energy splitting due the spatial confinemenent in each well,
 this system is well approximated by a Hubbard model, i.e. 
 by taking into account nearest-neighbor hopping and
 on-site interaction.\\

\noindent {\bf Nesting}\\
Fermi surface with portions that are parallel. The vector that connects different parallel portions is called the nesting vector $\vec{Q}$.

\section*{\textcolor{NavyBlue}{I. Definition of the Subject and Its Importance}}
\addcontentsline{toc}{section}{I. Definition of the Subject and Its Importance}
The work presented in this article belongs to the recently emerging
 interface of atomic physics and condensed matter theory.
 One of the crucial connections between these fields is the fact that
  ultracold atom ensembles in optical lattices, 
 i.e. periodic potentials provided by standing waves
 of laser light, 
 are well described
 by Hubbard models, the quintessential 
 model of many-body theory. 
 Therefore, these experiments
 allow for the study of many-body effects in a well-defined
 and tunable environment.

The subject of this article is the study
 of quantum phases of ultracold atoms in optical lattices.
 The objective is to propose experimental configurations, such as what
 lattice geometry or which types of atoms to use, 
 for which unusual many-body effects can be found. 
 Besides the applicability to ultra-cold atom systems, and given the generic
 nature of the underlying models,
 the resulting phases are also of interest in solid state systems.

 Using techniques such as a numerical implementation
 of functional renormalization group equations and Luttinger
 liquid theory, we find the phase diagrams
 of various low-dimensional systems of different 
 geometry, and discuss how the various phases could be detected.

\section*{\textcolor{NavyBlue}{II. Introduction}}
\addcontentsline{toc}{section}{II. Introduction}
The technology of cooling and trapping atomic ensembles has been one of
 the most important developments in physics over the last decades.
 It has been a critical ingredient in creating Bose-Einstein condensates 
 \cite{anderson, davis},
 improving atomic clocks \cite{phillips}, and studying atomic properties 
\cite{lett, jones}.
  A new direction in this development 
 was the realization of the Mott insulator transition
 \cite{greiner02} with ultra-cold atoms, 
 which demonstrated that these systems can be used
 to create various types of quantum phases in a 
 tunable and well-defined environment. 
 The subsequent 
  progress that has been made in controlling and
manipulating ensembles of ultra-cold
atoms~\cite{stoferle04,mandel03a,mandel03b,kohl05}, was followed by a number
of experiments to create and study more and more sophisticated
many-body effects, such as fermionic superfluids \cite{greiner03, jochim03, 
 zwierlein03},
one-dimensional strongly correlated Fermi and Bose
systems~\cite{paredes04, kinoshita04, Moritz}, or noise correlations in
interacting atomic systems~\cite{altman04, mathey05, greiner05, foelling05} . These developments
established the notion of 'engineering' many-body states in a
tunable environment, i.e. manipulating ensembles of ultra-cold atoms
in optical lattices.

This article further explores this development.
 The first step of creating novel states of matter is to determine
 the phase diagram of the system under consideration.
 For this purpose we use Luttinger liquid theory
 for studying one-dimensional quantum systems
  and two-dimensional thermal systems,
 and functional renormalization group equations
 to study two-dimensional quantum systems, which are both sophisticated
 methods that generate a lot of insight into the physics of these
 systems.

This article 
 contains three main sections, which can be read independently of
 each other, 
 organized
 as follows: in Section III
 we first study
 the phase diagram of an incommensurate Bose-Fermi mixture
 in one dimension, which can be understood as a Luttinger liquid
 of polarons (see \cite{mathey04, mathey07a}). We then 
 broaden the scope of this
 study to include the effects of commensurate densities (see \cite{mathey06b}).
 In Section IV, 
 we study
 the phases of two coupled two-dimensional superfluids, and
 we propose how the phase-locking transition of such systems
 can be used to realize  the Kibble-Zurek mechanism, i.e. to create
 topological defects by ramping across a phase transition (see \cite{mathey07c}).
 In Section V, we use a numerical implementation
 of functional renormalization group equations
 to study the phase diagrams
 of Bose-Fermi mixtures in
 optical lattices in two dimensions.
 For both a square and a triangular lattice
 we find a rich structure of competing phases (see \cite{mathey06a, mathey07b, klironomos07}).


\section*{\textcolor{NavyBlue}{III. One-Dimensional Lattices}}\label{1D}
\addcontentsline{toc}{section}{III. One-Dimensional Lattices}
%
%
%
The theory of one-dimensional many-body systems has been a highly active
 and fascinating field of physics for many decades, the centerpiece of which
 is the notion of the Luttinger liquid \cite{giamarchi, gnt, solyom}.
 In this section we propose several systems that display 
 various features of Luttinger liquids, such as quasi-long range order,
 competing orders, and Kosterlitz-Thouless transitions due to commensurate
 densities, as will be explained.

Recent advances in controlling ultra cold atoms
 lead to  the realization
of truly one dimensional systems, and the
 study of many-body effects therein. Important benchmarks,
such as the Tonks-Girardeau gas \cite{paredes04,kinoshita04} and
the Mott transition in one dimension\cite{stoferle04}, have been achieved
by trapping bosonic atoms in tight tubes formed by an optical
lattice potential. Novel transport properties of
one dimensional lattice bosons have been studied using these
techniques\cite{fertig}.
More recently, a strongly interacting one dimensional Fermi gas
was realized using similar trapping methods\cite{Moritz}.
Interactions between the fermion atoms
were
controlled
by tuning a Feshbach resonance
in these experiments.
On the theory side, numerous proposals were given for realizing 
a variety of different phases in ultra cold Fermi systems
\cite{recati03,fuchs04, cazalilla05},
 Bose-Fermi mixtures\cite{cazalilla03,mathey04,mathey07a,sengupta05}, as well as 
 Bose-Bose mixtures\cite{isacsson05a, isacsson05b}.

In the first part of this section, we describe the
 phase diagram of an incommensurate 
 Bose-Fermi mixture, in the second part
 we consider the effect of  commensurate fillings.

\subsection*{\textcolor{NavyBlue}{Luttinger liquid of polarons in one-dimensional Bose-Fermi mixtures}}
\addcontentsline{toc}{subsection}{Luttinger liquid of polarons in one-dimensional Bose-Fermi mixtures}
In this section we  investigate one dimensional (1D) Bose-Fermi mixtures (BFM) 
 using bosonization 
\cite{haldane_bf,cazalilla}. 
The resulting quantum phases can be understood by introducing 
polarons, i.e. atoms of one species
surrounded by screening clouds of the other species. 
In our analysis the polarons emerge as
the most well-defined quasi-particles in the interacting system
 while quantum phases of the system arise from a 
competition of various 
ordering instabilities of such polarons. 
The phase diagrams we obtain 
show a remarkable similarity to 
the Luttinger liquid phase diagrams of 1D interacting electron systems
\cite{solyom,jvoit}, suggesting that 1D BFM may be understood as Luttinger liquids
of polarons. 

\begin{figure}
\centerline{\includegraphics[width=6cm]{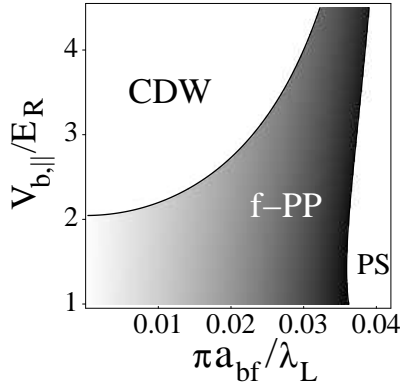}}
\caption{Phase diagram for a mixture of bosonic and spinless
fermionic atoms in a 1D optical lattice. 
Shading in the $f$-PP phase describes the strength
of the bosonic screening cloud ($2 \lambda$, see Eq.(\ref{polaron})) 
around a pair of fermions. $\lambda_L$ and 
$E_R$ are respectively the lattice period
and recoil energy.
Other parameters used for this figure are (see text for notations, 
 \cite{mathey07a} for details): 
$\nu_b=4$, $\nu_f=0.5$,
$V_{b,\perp}=V_{f,\perp}=20 E_R$, $V_{f,\|}=2 E_R$,  boson-boson 
scattering length $a_{bb}=0.01 \lambda_L$.
}
\label{phase_diagRP}
\end{figure}

\begin{figure}
\centerline{\includegraphics[width=6.5cm]{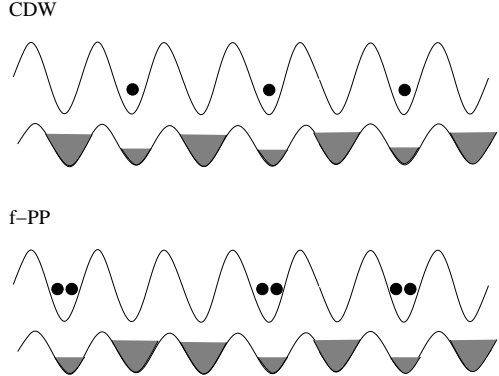}}
\caption{
%
Illustration of the two phases that occur 
in a BFM with spinless fermions, CDW and $f$-PP.
In the CDW phase the system develops a $2 k_f$
 density modulation in both the fermionic and the bosonic liquid.
 In the $f$-PP phase, 
 the fermions form polarons, indicated by the reduced bosonic density
 in their vicinity, that is, their polarization cloud.
 This polarization leads to an effective attractive interaction, which causes 
these fermionic polarons to pair up and form a 
 superfluid state. 
}
\label{CDW_fPP}
\end{figure}

To illustrate the results of this section, we show a typical
phase diagram for a BFM in an optical lattice in Fig. \ref{phase_diagRP}, 
as a function of experimentally controlled parameters.
 We consider two types of atoms, one fermionic and one bosonic, 
 moving in a lattice potential with the amplitude $V_{b,\|}$ 
 (see \cite{mathey07a}), and interacting via a short-ranged interaction
  characterized by 
 the scattering
length $a_{bf}$ between bosons and fermions.
 We use these parameters,
 the scattering
length $a_{bf}$ 
 and the strength
of the longitudinal optical lattice for bosonic atoms ($V_{b,\|}$)
\cite{endnote1}, as 
 tuning parameters in Fig. \ref{phase_diagRP}.
For relatively weak interactions and slow bosons (i.e. 
large $V_{b,\|}$) the system is in the charge-density-wave (CDW) phase, 
in which the densities of fermions and bosons have a periodic modulation
\cite{endnoteCDW}.
For very strong interactions the system is unstable to phase
separation (PS) \cite{cazalilla03,albus03,bijlsma00}. 
The two regimes are separated by a $p$-wave pairing phase of 
fermionic polarons  ($f$-PP).
Our analysis is carried out for the most promising system of atoms in an optical lattice. However, qualitative results should also apply to atoms in a tight 1D cigar-shaped magnetic trap \cite{goerlitz01}.
 A sketch of the two phases is shown in Fig. \ref{CDW_fPP}.

The essence of the bosonization procedure is to diagonalize the effective low-energy Hamiltonian, which allows for the exact calculation of all relevant correlation functions. The phase diagrams are determined by finding the order parameter which has the most divergent susceptibility \cite{solyom,jvoit}.
Bosonization  approach has been applied 
to BFM in Ref. \cite{cazalilla03}. 
However, that work did not consider the formation of polarons and,
as a result, did not describe most of the quantum phases discussed here.
The present system also has a close analogy to 1D electron-phonon 
systems discussed previously (see e.g.
Ref. \cite{voit_phonon}). A qualitative difference of the 
electron-phonon system is that the sound velocity is usually
much smaller than the Fermi velocity, whereas for a BFM the velocity
of the phonon modes (of the bosonic condensate) can be
larger than the Fermi velocity.
%
%
We also note that the 1-D $p$-wave superfluid we obtain here 
may be of relevance to a recent proposal for quantum computation \cite{kitaev}.

 We now give an overview over the, somewhat technical, derivation of this
 phase diagram, before we discuss issues concerning the
 experimental realization and detection of these phases, and conclude. 
 We consider a mixture of spinless fermionic
($f$) and bosonic ($b$) atoms. For a sufficiently strong optical potential the
microscopic Hamiltonian is given by a single band Hubbard model
\begin{eqnarray}
H &=& - \sum_{\langle ij\rangle} \left( t_b b_i^{\dagger} b_j 
+ t_f f_i^{\dagger} f_j \right)
-\sum_i\left(\mu_f n_{f,i}+\mu_b n_{b,i}\right)
+ \frac{U_{b}}{2} \sum_i  n_{b, i} (n_{b,i} - 1)
+U_{bf} \sum_i n_{b, i} n_{f, i},
\label{H_tot}
\end{eqnarray}
where $n_{b/f,i}$ are the boson/fermion density operators with $\mu_{b/f}$
being their chemical potentials.  The tunneling amplitudes $t_{f/b}$,
and the particle interactions $U_{b}$ and 
$U_{bf}$ can be expressed explicitly in terms of 
the $s$-wave scattering lengths,
the laser beam intensities and atomic masses \cite{jaksch98}. 
For simplicity we assume that the filling
fraction of fermions  $\nu_{f}\equiv\langle
n_{f,i}\rangle$ is not commensurate with the lattice
or with the filling fraction of bosons $\nu_{b}$.
Therefore, we can neglect lattice-assisted
backward/Umklapp scattering.
The  Fermi momentum and velocity are given by $k_f=\pi\nu_f$ 
and $v_f=2t_f\sin(k_f)$, respectively.

In Haldane's bosonization approach \cite{haldane_bf,cazalilla}
1D fermion and boson operators can be represented by
$f(x)=\left[\nu_f+\Pi_f\right]^{1/2}\sum_{m=-\infty}^\infty
e^{(2m+1)i\Theta_f}e^{i\Phi_f}$ and
$b(x)=\left[\nu_b+\Pi_b\right]^{1/2}\sum_{m=-\infty}^\infty
e^{2mi\Theta_b}e^{i\Phi_b}$,
where $x$ is a continuous coordinate that replaces the site index $i$.
The operators $\Pi_{f/b}(x)$ and $\Phi_{f/b}(x)$ are the bosonized
density and phase fluctuation operators.
The $\Theta_{f/b}(x)$ fields are given by $\Theta_{f/b}\equiv \pi \nu_{f/b} x + \pi \int^x dy \Pi_{f/b}(y)$.
The low-energy effective Hamiltonian thus can be written as:
\begin{eqnarray}
H_{\rm eff} &=& \sum_{\alpha=b,f}\frac{v_\alpha}{2}\int dx
\left[\frac{K_\alpha}{\pi}
\left(\partial_x\Phi_\alpha\right)^2+\frac{\pi}{K_\alpha}\Pi_\alpha^2\right]
 + U_{bf}\int dx \Pi_b\Pi_f
+\frac{2G}{2\pi}\int dx\left[\pi^2\Pi_f^2-(\partial_x\Phi_f)^2\right].
\nonumber\\
\label{H_eff}
\end{eqnarray}
where $v_b$ and $K_b$
are the phonon velocity and Luttinger exponent of the bosons 
and
$K_f=1$ for noninteracting fermion atoms. 
%
%
To obtain  the last term of $H_{\rm eff}$ we have integrated out the 
high energy ($2 k_f$) phonons 
within the instaneous approximation (i.e. assuming $v_b\gg v_f$). 
$G\equiv \frac{g_{2k_f}^2}{\omega_{2k_f}}$, where
$\omega_k$ is the (Bogoliubov) phonon energy dispersion 
\cite{ref_bogoliubiov} and
$g_k=U_{bf}\sqrt{\nu_b\,\varepsilon_{b,k}/2\pi\omega_k}$ 
is the fermion-phonon (FP) coupling vertex with $\varepsilon_{b,k}$
being the noninteracting boson band energy.
In the long wavelength limit we have a conventional
FP coupling $g_k=g |k|^{1/2}$ with $g\equiv U_{bf}\sqrt{K_b}/2\pi$.
The effective Hamiltonian, Eq. (\ref{H_eff}),
is quadratic and can be diagonalized \cite{Engelsberg}. The resulting two eigenmode velocities are
given by \cite{cazalilla03}
\begin{eqnarray}
v_{a, A}^2 & = & \frac{1}{2} (v_b^2 + \tilde{v}_f^2) \pm \frac{1}{2} 
\sqrt{(v_b^2 - \tilde{v}_f^2)^2 +  16 \tilde{g}^2  v_b \tilde{v}_f},
\label{new_velocity}
\end{eqnarray}
where $\tilde{v}_f\equiv(v_f^2-4G^2)^{1/2}$ and $\tilde{g}\equiv
g\,e^\theta$ with $e^\theta = 
((v_f - 2 G)/(v_f + 2 G))^{1/4}$.  
When the FP coupling $g$
becomes sufficiently strong the eigenmode velocity $v_A$ becomes imaginary, indicating an instability of the system. 
This instability corresponds to phase separation (global collapse) for 
positive (negative) $U_{bf}$\cite{cazalilla03}.

To understand the nature of the many-body state of BFM 
outside of the instability region
we analyze the long distance behavior of the correlation functions.
For the bare bosonic and fermionic particles we find $\langle
b(x)b^\dagger(0)\rangle \sim |x|^{-\frac{1}{2}K_\epsilon^{-1}}$ and
$\langle f(x)f^\dagger(0)\rangle \sim \cos(k_f x)
|x|^{-\frac{1}{2}(K_\beta+K_\gamma^{-1})}$
 \cite{Definitions}. 
To describe particles dressed
by the other species
we introduce the composite operators
\begin{eqnarray}
\tilde{f}(x) \equiv e^{-i\lambda\Phi_b(x)} f(x),
\ \ \ 
\tilde{b}(x) \equiv e^{-i\eta\Phi_f(x)} b(x),
\label{polaron}
\end{eqnarray}
with $\lambda$ and $\eta$ being some real numbers. The
correlation functions of these operators
are given by $\langle
\tilde{f}(x)\tilde{f}^\dagger(0)\rangle \sim \cos(k_f x)
|x|^{-\frac{1}{2}(K_\beta+\lambda^2 K_\epsilon^{-1} +K_\gamma^{-1}
-2\lambda K_{\gamma\epsilon}^{-1})}$ and $\langle
\tilde{b}(x)\tilde{b}^\dagger(0)\rangle \sim
|x|^{-\frac{1}{2}(K_\epsilon^{-1}+\eta^2 K_\gamma^{-1} -2\eta
K_{\gamma\epsilon}^{-1})}$ \cite{Definitions}.  
We observe that the exponents of the
correlation functions are maximized for
$\lambda_c=K_\epsilon/K_{\gamma\epsilon}$ and
$\eta_c=K_\gamma/K_{\gamma\epsilon}$. From now on we will use
Eq. (\ref{polaron}) with $\lambda_c$ and $\eta_c$ to construct polaronic
particles.
In the limit of weak
interactions we have $\lambda_c\to U_{bf}/U_b$ and
$\eta_c\to 2 U_{bf}/\pi v_b$. This result can be understood
by a simple density counting argument
that a fermionic polaron ($f$-polaron)
locally suppresses 
 (enhances)
a bosonic cloud by
$\lambda_c$ particles, whereas a bosonic polaron ($b$-polaron)
depletes 
 (enhances) 
the fermionic system by $\eta_c$ atoms 
 for positive (negative) $g$.
\begin{figure}
\centerline{\includegraphics[width=9.0cm]{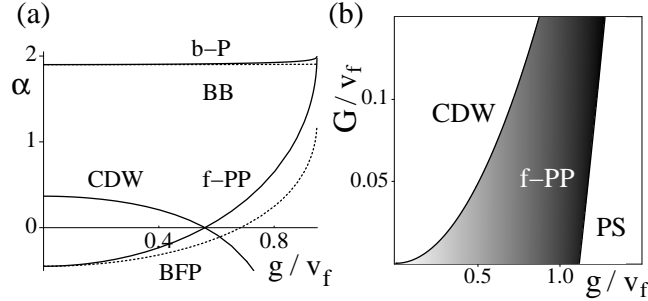}}
\caption{Ground state of a BFM with spinless fermions.
(a) Scaling exponents of different order parameters (see the text).
Parameters are chosen to be $v_b/v_f = 3$, $K_b = 5$ and $G/v_f = 0.1$.
(b) Global phase diagram for $v_b/v_f=5$
and $K_b=10$. Shading density indicates the strength of the screening
clouds of a polaron pair, $2\lambda_c$.}
\label{phase_diag_alpha}
\end{figure}

The polaronic operators defined in Eq. (\ref{polaron}) 
 can also be introduced via
 the canonical polaron transformation (CPT), 
which is often used in polaron theory \cite{mahan, alexandrov}. 
The CPT operator is given by
$U=e^{-i\lambda\sum_{{k}\neq 0}\left(
F_{k}\beta_{k}\rho_{k}^\dagger+{\rm h.c.}\right)}$, where $\beta_{k}$
is the phonon annihilation operator, $\rho_{k}$ is the fermion density
operator, $F_{k}$ is some function of wavevector $k$, and $\lambda$
specifies the strength of the phonon dressing. When applied to a
fermion operator, the CPT transforms it to a polaron
operator, $U^{-1} f(x)
U^{}=f(x)\exp\left[-i\lambda \sum_{{k}\neq 0}
\left(F_{k}\beta_{k} e^{-i{k}\cdot{x}}+{\rm h.c.}  \right)\right]$
\cite{mahan, alexandrov}, which is the same as Eq.(\ref{polaron}), provided
that one takes $F_k=\sqrt{\frac{2\pi}{K_b|k|L}}{\rm
sgn}(k)$. (Note that in 1D fermionic systems
density operators correspond to Luttinger bosons.) 
We note, however, unlike in ordinary polaron theory, where 
further approximations after the CPT have to be made \cite{mahan, alexandrov}, 
in the 1D BFM system we consider here, the full low
energy quantum fluctuations have been included via bosonization method
and exact diagonalization of the resulting Hamiltonian Eq. (\ref{H_eff}).
This allows for an essentially exact determination of the polarization parameter $\lambda$.


Now we study the many-body ground state phase diagram of a 1D BFM,
which is characterized by
specifying the order parameters that have the slowest long distance decay
of the correlation functions \cite{solyom,jvoit}. 
Two types of ordering were found to occur: $2 k_f$-ordering due 
to a Peierls-type instability and $f$-polaron pairing due to their 
effective attractive interactions induced by the screening clouds.
For the $2k_f$ CDW order parameter, $O_{CDW}=f^\dagger_{L} f_{R}$, 
we find $\alpha_{CDW}= 2-2K_\beta$,
and for the $f$-polaron pairing field,
$O_{f-PP}=\tilde{f}_{L}\tilde{f}_{R}$, we obtain
$\alpha_{f-PP}=2-2\left[\lambda^2_c K_\epsilon^{-1}
+K_\gamma^{-1}-2\lambda_c K_{\gamma\epsilon}^{-1}\right]$.
We did not include
polaron dressing in $O_{CDW}$, since this operator has no net
fermionic charge and the exponent
of $O_{CDW}$ does not change if we replace $f$ by $\tilde{f}$.
%
%
Scaling exponents shown in Fig. \ref{phase_diag_alpha}(a) demonstrate that
divergencies of the CDW and $f$-PP susceptibilities (corresponding to
positive $\alpha$) are mutually exclusive
and cover the entire phase diagram outside the PS regime. 
In the same figure, we also show the
scaling exponents calculated for bare fermion pairing ($O_{BFP}=f_Lf_R$),
bare boson condensate ($O_{BB}=b$), and $b$-polaron condensate ($O_{b-P}
=\tilde{b}$). 
It is easy to see that 
 the polaronic order parameters always
have larger exponents than their counterparts
constructed with bare atoms, showing the stability of polaronic
quasi-particle in a 1D BFM system.
Moreover, the necessity to consider $f$-polaron 
pairing instead of bare fermion pairing is further supported 
by considering the stability of superfluidity:
we introduce a single weak impurity potential in the 1D BFM
and determine its relevance 
by a
renormalization group (RG) calculation 
\cite{kane}. We find that the impurity potential is relevant within 
the CDW phase and irrelevant outside of it. This indicates 
that there should be a superfluid phase outside of the insulating CDW phase ,
which supports the existence of $f$-polaron pairing instead of 
bare fermion pairing
according to Fig. \ref {phase_diag_alpha}(a).

In Fig. \ref{phase_diag_alpha}(b) we show a global phase diagram of a BFM
considering the FP coupling ($g$) and effective fermion-fermion interaction
($G$) as independent variables. 
One can see that 
 the polaronic effects
and the associated pairing phase are important when FP coupling ($g$) is 
large, while the CDW phase dominates
when the effective fermion interaction ($G$) is increased.
This phase diagram is very
similar to what one finds for spinless electrons in Luttinger liquid
theory \cite{solyom,jvoit}, where CDW and pairing phase compete with each other
in the whole phase diagram. Therefore one can introduce a Luttinger
liquid of polarons to describe BFM in 1D systems.
The phase diagram in terms of experimentally 
controlled parameters was shown 
in Fig.~\ref{phase_diagRP}.
When considering finite temperature effects in a realistic experiment,
we note that the correlation function is cut-off by thermal 
correlation lengths, which are approximately 
given by $\xi \sim v_f /k_B T$. Therefore the zero temperature
ground states should appear when $\xi>L$ with $L$ being
the system size. This corresponds to a temperature regime of 
$1 \%$ of the Fermi temperature for systems of approximately 100 sites 
in the longitudinal direction.
%

Several approaches can be used to detect the quantum phases discussed above. 
First, in the CDW phase the fermion density modulation will
induce a $2k_f$ density wave
in the boson field in addition to the zero momentum 
condensation so that the CDW phase can be observed as interference peaks at momentum
$k=2k_f$ in a standard time-of-flight (TOF) measurement for bosons
\cite{endnoteCDW}.
Secondly, the polaron pairing phase can be
observed by measuring the noise correlation
of fermions in a TOF experiment as proposed in Ref. \cite{altman04}.
%
%
Thirdly, a laser stirring 
experiment \cite{onofrio99,raman99} can be used to probe 
the phase transition between the insulating (pinned by trap potential) 
CDW and the superfluid $f$-PP phase: one can use a laser beam
focused at the center of the cloud and 
stir such local potential to measure the response of the BFM.
If the system is in the pairing phase, the
laser beam can be moved through the system without
dissipation if only its velocity is slower than some critical
value \cite{onofrio99,raman99}. At the $f$-PP/CDW phase
boundary this critical velocity goes to zero, reflecting a transition
to the insulating (CDW) state. This scenario follows from 
the above described RG analysis of a single impurity potential \cite{kane}.
Finally a way to probe the PS boundary could be to 
measure the dipolar collective oscillations of the system,
generated by a sudden
displacement of the harmonic trap potential with respect to the lattice potential
\cite{maddaloni00,gensemer01,vichi99}. 
When the system is near the PS boundary, fermion-boson interaction 
will strongly reduce the frequency of the dipolar mode.

In summary, we used bosonization to
investigate the quantum phases of 1D mixtures of bosonic and 
fermionic atoms involving spinless
 fermions. 
The phase diagram that we found can be understood 
in terms of a Luttinger 
liquid of polarons.
We also described several experimental techniques for
probing these quantum phases. 

\subsection*{\textcolor{NavyBlue}{Commensurate mixtures of ultra-cold atoms in one dimension}}
\addcontentsline{toc}{subsection}{Commensurate mixtures of ultra-cold atoms in one dimension}
In this section we explore the behavior of ultracold atomic mixtures, confined 
to one-dimensional (1D) 
motion in an optical lattice, that exhibit different types of 
commensurability, by which
 we mean that the atomic densities and/or the inverse 
 lattice spacing have an integer ratio.
Commensurable fillings arise naturally in many ultracold atom systems, because 
the external trap potential approximately corresponds to a sweep of the 
chemical potential through the phase diagram, and therefore passes through points 
of commensurability.
At these
points the system can
 develop an energy gap, which fixes
 the density commensurability over a spatially extended volume.
%
%
%
%
%
%
%
%
%
%
This was demonstrated in 
 the celebrated Mott insulator 
experiment by Greiner et al.\cite{greiner02}, 
 where Mott phases with integer filling 
 occurred in shell-shaped regions in the atom trap.
These gapped phases gave rise to the  
 well-known signature in the time-of-flight images\cite{trap},
 and triggered the endeavor of `engineering' many-body states
 in optical lattices.
Further examples include the recently created density-imbalanced
 fermion mixtures \cite{partridge05,zwierlein05} in which the development of
 a balanced, i.e. commensurate,
  mixture at the center of the trap is observed.

In 1D, this phenomenon is of particular importance,
 because it is the only effect that can lead to the opening of
 a gap, for a system with short-range interactions. 
 In contrast to higher dimensional systems, where, for instance, pairing
 can lead to a state with an energy gap,
 in 1D  only discrete symmetries
 can be broken, due to the importance of fluctuations.
Orders that correspond to a
 continuous symmetry can, at most, develop quasi long range order (QLRO),
 which refers to a state in which an order parameter $O(x)$
 has a correlation function with algebraic scaling, 
$\lav O(x) O(0)\rav \sim |x|^{-(2-\alpha)}$, with
 a positive scaling exponent $\alpha$.

Due to its importance in solid state physics, the most thoroughly
 studied commensurate 1D system is the SU(2) symmetric
 system of spin-1/2 fermions.
 This system develops a spin gap for attractive 
 interaction and remains gapless 
 for repulsive interaction, as
 can be seen from a second order RG calculation.
 However, the assumed symmetry between the
 two internal spin states, which is natural in solid state systems,
 does not generically occur 
in Fermi-Fermi mixtures (FFMs) 
of ultra-cold atoms, where
 the `spin' states are in fact different 
 hyperfine states of the atoms. 
 An analysis of the generic system 
 is therefore highly called for.
 Furthermore, we will extend this analysis
 to both Bose-Fermi (BFMs) and Bose-Bose mixtures (BBMs), as well
 as to the dual commensurability, in which the charge field, and not
 the spin field, exhibits commensurate filling, as will be explained
 below.

\begin{figure}
\centerline{\includegraphics[width=9.2cm]{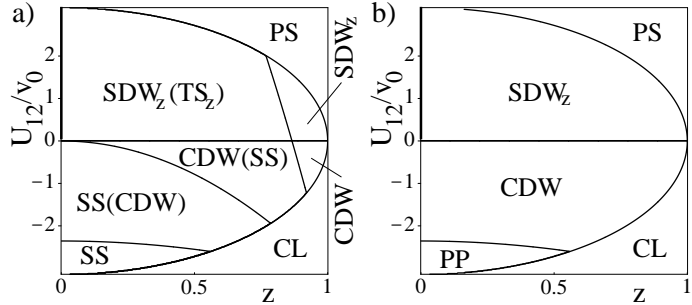}}
\caption{\label{Spin_gap_FF}
a) 
Phase diagram of a commensurate FFM or a
 BBM of
 hardcore bosons (with the replacement $TS_z\ra SS$), 
b) phase diagram of a BFM with hardcore bosons, 
in terms of the interaction $U_{12}$
and the parameter $z=|v_1-v_2|/(v_1+v_2)$.
For both attractive  and repulsive 
 interactions a spin gap opens, except for
 $z=0$ and positive interaction. 
In the attractive regime, a FFM or a BBM shows
 either singlet pairing or CDW order, or a coexistence of these phases,
 a BFM shows either CDW order or polaron pairing.
 For repulsive interaction
 all mixtures show SDW ordering, with FFMs and BBMs
 showing subdominant triplet or singlet pairing, respectively,
  for a large range of $z$.
In the gapless
 regime, a FFM shows degenerate 
SDW and CDW order, a BFM shows CDW order for the fermions
 and SF for the bosons, and a BBM shows SF with subdominant 
 CDW, i.e. supersolid behavior.
For very large positive values of $U_{12}$ the system undergoes
phase separation (PS); for very large negative values it collapses (CL).
}
\end{figure}

\begin{figure}
\centerline{\includegraphics[width=9.2cm]{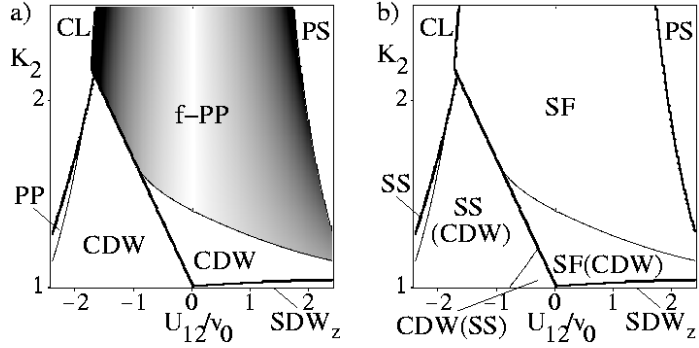}}
\caption{\label{Spin_gap_BF}
a) Phase diagram of a BFM,
 b) phase diagram of a BBM with the first 
species being in the hardcore limit,
 in 
terms of $U_{12}$, and 
the Luttinger parameter of the second species ($K_2$), at
 the fixed 
 velocity ratio $|v_1-v_2|/(v_1+v_2)=0.5$.
For large repulsive interaction the system undergoes phase separation (PS), for large attractive
interaction the system collapses (CL).
In the regime below the thick line the system opens
 a gap, i.e. if species 2 is close to the hardcore limit.
 However, for larger values of $K_2$,  
the gapless phase is restored.
 Close to the
 transition, the properties of the fermions, respectively hardcore bosons,
 are still affected by the RG flow, leading
 to CDW order for the fermions
 and to supersolid behavior for the bosons.
}
\end{figure}

The main results of this section are the phase
 diagrams shown in Fig. \ref{Spin_gap_FF} and \ref{Spin_gap_BF}.
 We find that both attractive and repulsive interactions
 can open an energy gap.
 For FFMs the entire phase diagram is gapped, except for
 the repulsive SU(2) symmetric regime (cp. \cite{cazalilla05}), for
 BFMs or BBMs the bosonic liquid(s) need(s) to be close to the
 hardcore limit, otherwise the system remains gapless.
Furthermore, we find a rich structure of quasi-phases,
 including charge and spin density wave order (CDW, SDW), 
 singlet and triplet pairing (SS, TS), polaron pairing \cite{mathey04,
 mathey07a},
 and a supersolid phase, which is the first example 
 of a supersolid phase in 1D.
These results are derived within a Luttinger liquid (LL) 
 description, which treats bosonic and fermionic liquids on equal footing.
%
%
%
%
%
%


We will now classify the
 types of commensurability that can occur in
 a system with short-ranged density-density
 interaction.
 We consider Haldane's representation \cite{haldane_bf, cazalilla}
of the densities for the two species:
\bea
n_{1/2} & = & [\nu_{1/2} + \Pi_{1/2}] \sum_{m} e^{2 m i\Theta_{1/2}}
\eea
$\nu_1$ and $\nu_2$ are the densities of the two liquids,
$\Pi_{1/2}(x)$ are the low-k parts (i.e. $k\ll 1/\nu$) of the density fluctuations; 
 the fields  
$\Theta_{1/2}(x)$ are given by 
 $\Theta_{1/2}(x) 
= \pi \nu_{1/2} x + \theta_{1/2}(x)$, 
with $\theta_{1/2}(x)=\pi \int^x dy \Pi_{1/2}(y)$.
%
%
%
 These expressions hold for both bosons and fermions.
If we use this representation in a density-density 
 interaction 
term $U_{12}\int dx n_1(x)n_2(x)$,
we generate to lowest order a term of the shape $U_{12}\int dx \Pi_1(x)\Pi_2(x)$, 
 but in addition an infinite number of nonlinear terms, corresponding
to all harmonics in the representation.
However, only the terms for which the linear terms ($2 \pi m_{1/2} \nu_{1/2}x$) cancel,
can drive a phase transition.
For a continuous system this happens for $m_1 \nu_1 - m_2\nu_2=0$,
whereas for a system on a lattice we have the condition $m_1 \nu_1 - m_2\nu_2=m_3$, 
where $m_1$,$m_2$ and $m_3$ are integer numbers.
%
%
In general, higher integer numbers correspond to terms
 that are less relevant, because the scaling dimension
 of the non-linear term scales quadratically with these integers.
 We are therefore lead to consider
 small integer ratios between
 the fillings and/or the lattice if present.
In \cite{mathey07a}, we considered
 two cases of commensurabilities: a Mott insulator transition
 coupled to an incommensurate liquid,
 and a fermionic liquid at
 half-filling coupled to an incommensurate bosonic liquid. 
%
In both cases
 the
 commensurability occurs between  one species and the
 lattice, but does not involve the second species.
%
%
%
 Here,
 we 
 consider the 
 two most relevant, i.e. lowest order,  
cases which exhibit a commensurability that 
 involves both species. 
 The first case is the case 
 of equal filling $\nu_1=\nu_2$, 
 the second is the case 
 of the total density being unity, i.e. $\nu_1+\nu_2=1$,
 where the densities $\nu_{1}$ and $\nu_2$ themselves are incommensurate.
The first case can drive the system to a spin-gapped state, the
 second to a charge gapped state.
 We will determine in which parameter
 regime these transitions occur,
 and what type of QRLO the
 system exhibits in the
 vicinity of the transition. 
These two cases can be mapped onto each other via a dual mapping, which
enables us to study only one
 case and then infer the results for the
 second by using this mapping.
We will 
write out our discussion
 for the case of equal filling
 and merely state 
the corresponding results for 
 complementary filling.

The action of a two-species mixture
with equal filling
in bosonized form is given by:
\bea\label{S}
S & = & S_{0,1} + S_{0,2} + S_{12} +S_{int}.
\eea
The terms $S_{0,j}$, with $j=1,2$, are given by
\bea
S_{0, j} & = & \frac{1}{2\pi K_{j}} \int d^2r \Big( \frac{1}{v_{j}}(\partial_{\tau} \theta_{j})^2 +  v_{j} (\partial_x \theta_{j})^2 \Big)
\eea
Each of the two types of atoms, regardless of being bosonic or fermionic,
are characterized by a Luttinger parameter $K_{1/2}$ and a velocity $v_{1/2}$.
Here we integrate over ${\bf r}=(v_0 \tau, x)$, 
 where we defined the
 energy scale $v_0=(v_1+v_2)/2$.
%
%
%
%
The term $S_{12}$ describes the acoustic coupling between the two species, and is bilinear:
\bea
S_{12} & = &  \frac{U_{12}}{\pi^2} \int d^2r \partial_x \theta_1 \partial_x \theta_2 
 + \frac{V_{12}}{\pi^2} \int d^2r \partial_\tau \theta_1 \partial_\tau \theta_2.
\eea
The second term is created during the RG flow; its prefactor therefore has the initial value
$V_{12}(0)=0$.
We define $S_0 =  S_{0,1} + S_{0,2} + S_{12}$, which is
 the diagonalizable part of the action. 
%
%
%
$S_{int}$ 
 corresponds 
to the non-linear coupling between the two liquids,
which we study within an RG approach:
\bea
S_{int} & = &  \frac{2 g_{12}}{(2 \pi \alpha)^2} \int d^2r  
\cos(2 \theta_1 - 2\theta_2).
\label{Sint}
\eea
%
%
%
%
%
%
%
This 
 bosonized 
description 
  applies to a BBM, a
 BFM, and a FFM.
Depending on which of these mixtures 
we want to describe we either 
construct bosonic or fermionic operators 
according to Haldane's contruction \cite{haldane_bf,cazalilla}:
\bea
f/b & = & [\nu_0 + \Pi]^{1/2} \sum_{m\, odd / even} e^{m i\Theta} e^{i\Phi}.
\eea
$\nu_0$ is the zero-mode of the density,
$\Phi(x)$ is the phase field, which is the conjugate field of the density fluctuations $\Pi(x)$.
The action for a mixture with complementary
 filling, $\nu_1+\nu_2 =1$, is of the form $S_0+S'_{int}$, where
 the interaction $S'_{int}$ is given by:
\bea
S'_{int} & = &  \frac{2 g_{12}}{(2 \pi \alpha)^2} \int d^2r  
\cos(2 \theta_1 + 2\theta_2).
\label{Sint'}
\eea
To map the action in Eq. (\ref{S}) onto
 this system we use
 the mapping: $\theta_2\ra -\theta_2$, $\phi_2\ra -\phi_2$, and 
$g_{12}\ra -g_{12}$, which evidently
 maps a mixture with complementary filling and attractive (repulsive)
 interaction and onto a mixture with equal
 filling with repulsive (attractive) interaction.

 To study the action given in Eq. (\ref{S}), 
we perform an RG calculation along the lines of the
treatment of the sine-Gordon model in \cite{kogut79,gnt}.
In our model, 
 a crucial modification
 arises: the linear combination  
$\theta_1 - \theta_2$, that appears
 in the non-linear term,  is not proportional to an eigenmode
of $S_0$, and therefore the RG flow does not affect only  one
 separate sector of the system, as in an
SU(2)-symmetric system.
%
The RG scheme that we use here
 proceeds as follows:
First, we diagonalize $S_0$ through the transformation 
 (see \cite{mathey06b})
%
%
%
%
$\theta_1  =  B_1 \tilde{\theta}_1 + B_2\tilde{\theta}_{2}$,  
 and
$\theta_2  =  D_1 \tilde{\theta}_1 + D_2\tilde{\theta}_{2}$,
%
%
%
%
%
%
%
%
%
%
%
%
%
%
%
%
%
%
 where $B_{1/2}$ and $D_{1/2}$ are some coefficients,
 and
 $\tilde{\theta}_{1/2}$ are the eigenmode
 fields with velocities
  $\tilde{v}_{1/2}$.
%
%
%
%
%
%
%
%
%
%
%
%
Now we introduce an energy cut-off $\Lambda$ on
 $\tilde{\theta}_{1/2}$ 
according to
$\omega^2/\tilde{v}_{1/2} + \tilde{v}_{1/2}k^2 < \Lambda^2$.
We shift this cut-off by an amount $d\Lambda$,
and correct for this shift up to second order in $g_{12}$.
At first order, only $g_{12}$
  is affected, its
flow equation is given by:
\bea
\frac{d g_{12}}{d l} & = & \Big(2 - K_1 - K_2 - 
\frac{2}{\pi}\frac{U_{12} + V_{12} v_1 v_2}{v_1 + v_2}\Big) 
g_{12}, \label{RG_g12}
\eea
%
%
%
 with $dl=d\Lambda/\Lambda$. 
At second order several terms are created that are 
quadratic in the original fields $\theta_1$ and $\theta_2$.
We undo the diagonalization, and absorb
these terms into the parameters of the action, which
 concludes the RG step.
 By iterating this procedure we obtain these flow equations
 at second order
in $g_{12}$:
\bea
\frac{d K_{1/2}}{d l} & = & - \frac{g_{12}^2}{16\pi^2} 
\Big(2+\Big(\frac{v_2}{v_1}+\frac{v_1}{v_2}\Big)\Big)\label{RG_K1}\\
\frac{d v_1}{d l} & = &   v_1\frac{g_{12}^2}{16\pi^2} 
\Big(\frac{v_2}{v_1}-\frac{v_1}{v_2}\Big)\label{RG_v1}\\
\frac{d v_2}{d l} & = & v_2 \frac{g_{12}^2}{16\pi^2} 
\Big(\frac{v_1}{v_2}-\frac{v_2}{v_1}\Big)\label{RG_v2}\\
\frac{d U_{12}}{d l} & = & - \frac{g_{12}^2}{8\pi} (v_1+v_2)\label{RG_U12}\\
\frac{d V_{12}}{d l} & = & - \frac{g_{12}^2}{8\pi} (1/v_1+1/v_2)\label{RG_V12}
\eea
%
%
%
%
%
%
%
%
%
%
%
%
%
%
%
%
%

%
%
The system of differential equations, Eqns. (\ref{RG_g12}) 
to (\ref{RG_V12}),
 can show two types of qualitative behavior:
 The coefficient $g_{12}$ of the non-linear term (\ref{Sint}) can
 either flow to zero, i.e. $S_{int}$ is irrelevant, or it diverges,
 leading to the formation of an energy gap.
 In the first case, the system flows to
 a fixed point that is described
 by a renormalized diagonalizable
  action of the type $S_0$, from which the quasi-phases can be
 determined.

%
%
%

 When $S_{int}$ is relevant,
%
 we 
introduce the fields \cite{jvoit}
%
%
%
$\theta_{\rho/\sigma} = \frac{1}{\sqrt{2}}(\theta_1 \pm \theta_2)$,
%
%
%
which define the charge and the spin sector of the system.
 In this regime, 
 these sectors 
 decouple.
%
Each of the
 two sectors is characterized by a Luttinger parameter
and a velocity, $K_{\rho/\sigma}$ and $v_{\rho/\sigma}$,
 which are related to the original parameters
 in $S_0$ in a
straightforward
 way. 
Using the numerical solution of the flow equations, 
we find that $K_\sigma\rightarrow 0$, as can be expected for
an ordering of the nature of a spin gap, leaving $K_\rho$
the only parameter characterizing the QLRO in this phase.
%
%
%
%
%
%
%
%
%
%
%
%
%
%

In order to determine
 the QLRO in the system we will determine
 the scaling
 exponents of various order parameters.
 The order parameter with the largest positive scaling exponent
 shows the dominant order, whereas other orders with positive
 exponent are subdominant.

We will now apply this procedure to the different types of mixtures.
For a FFM 
 we find that the system always develops a
 gap, with the exception of the repulsive SU(2) symmetric regime 
 (cp. \cite{cazalilla05}).
 To determine the QLRO 
we introduce the following operators \cite{jvoit, giamarchi}:
$O_{SS} =  \sum_{\sigma, \sigma'} \tilde{\sigma} f_{R,\sigma} \delta_{\sigma,\sigma'} 
f_{L, 3-\sigma'}$, 
$O_{TS}^a =  \sum_{\sigma, \sigma'} \tilde{\sigma} f_{R,\sigma} \sigma^a_{\sigma,\sigma'}
 f_{L, 3-\sigma'}$, 
$O_{CDW} = \sum_{\sigma, \sigma'} f_{R,\sigma}^\dagger \delta_{\sigma,\sigma'} 
f_{L,\sigma'}$, and
 $O_{SDW}^a =  \sum_{\sigma, \sigma'} \tilde{\sigma} f_{R,\sigma}^\dagger 
\sigma^a_{\sigma,\sigma'}
 f_{L,\sigma'}$, 
 with $\sigma, \sigma'=1,2$, $\tilde{\sigma}=3-2\sigma$, and $a=x,y,z$. 
In the gapless SU(2) symmetric regime, 
 both CDW and SDW show QLRO, with both scaling exponents of the
 form $\alpha_{SDW/CDW}=1-K_\rho$ \cite{jvoit}, which
 shows that these
 orders are algebraically degenerate.
Within
 the gapped regime
 the scaling exponents of 
these operators are given by $\alpha_{SS,TS_z}= 2 - K_\rho^{-1}$ and
$\alpha_{CDW,SDW_z} = 2 - K_\rho$. 
As discussed in \cite{giamarchi}, the
 sign of $g_{12}$ determines
 whether CDW or SDW$_z$, and SS or TS$_z$ appears.
 In Fig. \ref{Spin_gap_FF} a), 
we show the phase diagram based on these results.
 In addition to these phases we indicate the appearance
 of the Wentzel-Bardeen instability, shown as phase separation
 for repulsive
 interaction and collapse for attractive interaction.

We will now use the dual mapping
 to obtain the
 phase diagram of a FFM with complementary
 filling from Fig. \ref{Spin_gap_FF} a). 
 Under this mapping, the attractive and repulsive regimes
 are exchanged with the following replacements: 
 $CDW\ra SDW_z$, $SDW_z\ra CDW$, $SS,TS_z\ra SDW$, and $SDW\ra SS$.
 Note that the gapless regime
 is now on the attractive side, 
  with degenerate CDW and SS pairing.

%
%
%
%
%
%
%

For BBMs we proceed in the
 same way as for FFMs. 
We introduce the following set of order parameters:
 $O_{CDW}=b_1^\dagger b_1 +b^\dagger_2 b_2$, 
$O_{SS}=b_1 b_2$,
  $O_{SDW_z}=b_1^\dagger b_1 -b^\dagger_2 b_2$, 
  $O_{SDW_x}=b_1^\dagger b_2 +b^\dagger_2 b_1$,
  $O_{SDW_y}=-i (b_1^\dagger b_2 -b^\dagger_2 b_1)$, and
 in addition the superfluid (SF) order
 parameters $b_1$ and $b_2$.
 In  Fig. \ref{Spin_gap_FF} a) 
 we show the phase
 diagram of a mixture of a BBM of hardcore bosons, which
 is almost identical to the one of a FFM.
The phase diagram of the mixture with complementary
 filling, as obtained from the dual mapping, is
 also of the same form as its fermionic equivalent,
 with the
 exception of the gapless regime, in which BBMs show
 supersolid behavior (coexistence
 of SF and CDW order), and with the replacement $TS_z\ra SS$.

In Fig. \ref{Spin_gap_BF} b), we  
show the phase diagram of 
 a mixture of hardcore bosons (species 1) and bosons in the intermediate
 to hardcore regime (species 2).
 If species 2 is sufficiently far
 away from the hardcore limit, the system remains gapless.
 However, in the vicinity
 of the transition the scaling exponents of the liquids
 are affected by the RG flow. As indicated,
 the effective scaling exponent
 of the hardcore bosons 
  is renormalized to a value that is smaller than 1, and therefore
 we find both SF and CDW order, i.e. supersolid behavior.
 The phase diagram of the dual mixture is of the following form:
  the attractive and the repulsive regime are exchanged,  
 and in the gapped phase we again have the 
 mapping: 
 $CDW\ra SDW_z$, $SDW_z\ra CDW$, $SS\ra SDW$, and $SDW\ra SS$.
 The gapless regime is unaffected.


For a BFM 
 we find that the order parameters $O_{CDW}$, $O_{SDW_z}$, 
 $O_{f-PP}=f_R f_L e^{-2i\lambda \Phi_b}$ \cite{mathey04, mathey07a},
and $b$ can develop QLRO in the gapless regime. 
%
In the gapped regime, the order parameters
$O_{PP}\equiv f_R b f_L b$ 
 and $O_{PP'}\equiv f_R b^\dagger f_L b^\dagger$, in addition to $O_{CDW}$,
show 
  QLRO.
 ($O_{PP/PP'}$ are special cases of the polaron pairing operators discussed
 in \cite{mathey04,mathey07a}.) 
In Fig. \ref{Spin_gap_FF} b) we show
 the phase diagram
 of a BFM with hardcore bosons, and
  in Fig. \ref{Spin_gap_BF} a), 
 we vary the Luttinger parameter
 of the bosons.
 In both the gapless phase and the gapped phase, we find that
 CDW and $f$-PP or PP, respectively, are mutually exclusive
 and cover the entire phase diagram, cp. \cite{mathey04,mathey07a}.
The dual mapping 
 again maps attractive and repulsive regimes onto each other. Within
 the
 gapped phase we find the mapping
 $CDW\ra SDW_z$, $SDW_z\ra CDW$, and $PP\ra PP'$, the 
gapless regime is unaffected.

%
%
%
%

 %


%
%
%
%
%

Before we conclude,
 we discuss how these
 predictions 
 could be measured experimentally.
CDW order
 will create additional peaks in TOF images, corresponding to
 a wavevector $Q=2k_f$.
 As demonstrated and pointed out in \cite{greiner05,foelling05,altman04,mathey05},
 the noise in TOF images
 allows to identify
   the different regimes
 of both gapped and gapless phases.
As discussed in \cite{mathey04,mathey07a},
 a laser stirring experiment could determine
 the onset of CDW order for fermions, or the supersolid regime
 for bosons.
 RF spectroscopy \cite{chin} can be used
 to determine
 the presence and the size
 of an energy gap.

In conclusion, we have studied mixtures of ultra-cold
 atoms in 1D with commensurate filling.
We used a Luttinger liquid description which enables us to
  study FFMs, BFMs, and BBMs in a single
 approach. 
We find that FFMs are generically gapped for both attractive and repulsive interactions,
 whereas for BFMs and BBMs  the bosons need to be close
 to the hardcore limit.  
 We find a rich structure of quasi-phases in the vicinity 
 of these transitions, in particular a supersolid
 phase for BBMs, that occurs 
  close to the hardcore limit. 
Experimental methods to detect the predictions were also discussed.

\section*{\textcolor{NavyBlue} {IV. Phase-locking transition of coupled low-dimensional superfluids}}\label{PL}
\addcontentsline{toc}{section}{IV. Phase-locking transition of coupled low-dimensional superfluids}

Most phase transitions that have been realized in ultra-cold
 atom systems are generic first or second order transitions.
 However, the paradigm of phase transitions in two dimensions
 at finite temperature is of a more intricate type, 
 a Kosterlitz-Thouless transition, which is characterized
 by a change of the functional form of the correlation function
 of the order parameter, from algebraic decay to exponential decay.
  In an intriguing new development in studying
low-dimensional strongly correlated systems,  such a 
   Kosterlitz-Thouless
(KT) transition~\cite{chaikin-lubensky} was indeed
 realized and observed~\cite{zoran}. In this experiment the
interference amplitude between
 two
 independent two-dimensional
(2D) Bose systems was studied as a function of temperature. This
analysis revealed the jump in the superfluid stiffness (see also
Ref.~\cite{pad}) and the emergence of unpaired isolated
vortices as they crossed the phase transition.

 The other focus of this section, the physics of ramping across
 a phase transition, is also triggered by 
 a recent experiment:  Sadler {\em et. al.}
observed spontaneous generation of topological defects in the spinor
condensate after a sudden quench (i.e. a rapid, non-adiabatic
 ramp) through a quantum phase
transition~\cite{kurn}. A similar experiment in a double-layer
system was reported in Ref.~\cite{scherer}. The topological
defects are generated~\cite{kibble} at a density which is related to
the rate at which the transition is crossed~\cite{zurek}. Later it
was argued that the dependence of the number of such defects on the
swipe rate across a quantum critical point can be used as a probe of
the critical exponents characterizing the phase
transition~\cite{kbz}. 
 This Kibble-Zurek (KZ) mechanism 
 was originally considered as an early universe scenario 
 creating cosmic strings, which 
 would serve as an 
 ingredient for the formation of galaxies \cite{kibbleremark}. 
 Cold atom systems appear
to be a very  
 suitable laboratory for performing such
``cosmological experiments'', since these systems are
highly tunable and well isolated from the environment. So far the
experiments and the theoretical proposals addressed the KZ 
scenario across a quantum phase transition. The main reason is that
it is generally hard to cool such systems sufficiently fast to
observe non-equilibrium effects. In this work 
 we provide 
 an example of a
particular system 
 where 
this difficulty can be easily
overcome by quenching the transition temperature $T_c$ instead of
$T$. Thus the relevant ratio $T/T_c$ can be tuned with an arbitrary
rate and the KZ mechanism can be observed. 
 Specifically, we
examine a system of two superfluids (SF):
As we show below, by turning on tunneling 
 between the two systems
 the transition
temperature
  increases rapidly, 
 and the system attempts to create long-range
order (LRO).
 However, in this process, defects in the SF
phase are created, which develop into long-lived vortex-anti-vortex
pairs or in finite system unbalanced population between vortices and
anti-vortices. We note that because the systems are isolated and
there is no external heat bath, the temperature itself also changes
due to the quench. However, the long-wavelength fluctuations
relevant for the KT transition are only a small subset of all
degrees of freedom, majority of which are only weakly affected by
small inter-layer tunneling. So we believe that the change of the
$T_{c}$ is the main effect of the quench.

\begin{figure}
\centerline{\includegraphics[width=6.0cm]{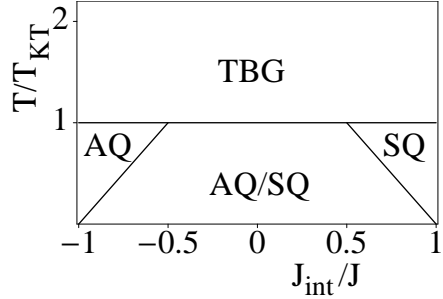}}
\caption{\label{PD_Jint} Phase diagrams of two 2D SFs, coupled
 through a term of the form ${\mathcal S}_{12}$, Eq. (\ref{S12}), 
in terms of $J_{int}/J$ and $T/T_{KT}$. 
 For low temperatures we find antisymmetric quasi-order (AQ) and/or 
 symmetric quasi-order (SQ), which either simultaneously undergo a KT
 transition due to single vortices  (AQ/SQ to thermal Bose gas (TBG) phase), 
 or individually due to correlated vortex
 pairs: symmetric (anti-symmetric) vortex pairs drive the AQ/SQ to AQ (SQ) transition.
}
\end{figure}

In this section we consider two SFs coupled via tunneling and/or
interactions. In the experiments the hopping or tunneling rate
between two systems can be tuned to a high precision
\cite{paredes04,zoran,joerg,zoran1}. Interactions between the atoms in
different systems can either be realized in ensembles of polar
molecules or by using mixtures of two hyperfine states, where the
tunneling rate is controlled by an infrared light
source~\cite{phase}, which induces 
 spin-flipping between the hyperfine states. 
 In this case the atoms in different states
naturally interact with each other since they are not physically
separated in space. 
 The main results of our analysis are
the phase diagrams of coupled SFs in Fig.~\ref{PD_Jint} and
 Fig. \ref{pdint}, the behavior of $T_{c}$ and the energy gap
shown in Fig.~\ref{PD}, as well as
 the proposal of realizing the KZ mechanism by
 switching on the tunneling between two SFs.
%
%
%
%
%
%
\subsection*{\textcolor{NavyBlue}{2D superfluids}}
\addcontentsline{toc}{subsection}{2D superfluids}
 In this section we consider 
two 2D SFs, each characterized
by a KT temperature $T_{KT}$. 
 We write the bosonic
operators $b_{1/2}$ in the two layers in a phase-density
representation~\cite{chaikin-lubensky,gnt}, $b_{1/2}\sim
\sqrt{\rho_{1/2}} \exp(i \phi_{1/2})$, where $\rho_{1/2}$ are 
 the density operators of the two 
 systems, and 
 $\phi_{1/2}$ the phases. 
 The low-momentum fluctuations
of the phase fields are described by Gaussian contributions to the
Hamiltonian $\mathcal H_0$. Because of the formal analogy between
the quantum 1D and thermal 2D systems~\cite{giamarchi} we adopt
the quantum terminology throughout the paper and refer to the ratio
of the Hamiltonian and the temperature as the action. Then
\beq
{\mathcal S}_{0}\equiv \frac{{\mathcal H}_0}{T} = \frac{J}{2 T} \int d^2r
[(\nabla \phi_1)^2+(\nabla \phi_2)^2] \,.
\label{S0}
\eeq
The energy scale $J$ here is related to $T_{KT}$ by $J=2
T_{KT}/\pi$.
Besides these long-wavelength fluctuations, the system also contains
additional degrees of freedom, vortex-anti-vortex
pairs~\cite{chaikin-lubensky}. The corresponding term in the action
  is expressed through the dual fields $\theta_{1,2}$~\cite{gnt}:
\bea\label{S1}
{\mathcal S}_1 & = &  \frac{2 A_1}{T} \int \frac{d^2 r}{(2 \pi\alpha)^2}
[\cos(2\theta_1) + \cos(2\theta_2)] \, ,
\eea
where $\alpha$ is a short-distance cut-off of the size of the vortex
core, and 
 $A_1$ 
 is proportional to the
 single-vortex fugacity: $A_1\sim J\exp(-J/T)$,  
 where we assume 
 both SFs 
 to have the same effective parameters $J$ and
 $A_1$. 
Operators of the type
$\exp(2 i \theta)$ create kinks in the 
field $\phi$: $\exp(-2
i \theta(x))\phi(x')\exp(2 i \theta(x)) \sim \phi(x') + 2\pi \Theta(x
-x')$, $\Theta(x)$ being the step function, which 
 corresponds to
the effect of vortices in the original 2D problem 
 (Ref.~\cite{giamarchi}, p. 92).

In addition the two systems are coupled by a hopping term $\sim
t_\perp b_1^\dagger b_2 + h.c.$, which results in the following
contribution to the action:
\bea\label{Sperp}
{\mathcal S}_\perp & = &  \frac{2 J_\perp}{T} \int \frac{d^2 r}{(2 \pi\alpha)^2} \cos(\phi_1-\phi_2),
\eea
where the bare value of $J_\perp$ corresponds approximately to
$t_\perp \rho_0$. In principle, the hopping term is modified by the
vortex contributions, however, these corrections are always
irrelevant under renormalization group (RG). 

For most of the discussion in this paper
 we use the symmetric and anti-symmetric combinations of 
 $\phi_{1/2}$ and $\theta_{1/2}$:
\beq
\phi_{s/a} = (\phi_1 \pm \phi_2)/\sqrt{2} ,\quad
\theta_{s/a} = (\theta_1 \pm \theta_2)/\sqrt{2} \, .
\label{spincharge}
\eeq
 Written in 
 these fields, 
the term ${\mathcal S}_0$ in
 Eq. (\ref{S0}) is again a sum of Gaussian models, now in 
 the fields $\phi_s$ and $\phi_a$,
 with the same energy scale $J$.
 However, we will consider a broader class of actions,
 in which the energy scales of the symmetric and anti-symmetric sector
  differ. We include the following term in the action:
\bea\label{S12}
{\mathcal S}_{12} & = &  \frac{J_{{\rm int}}}{T} \int d^2r \nabla \phi_{1}\nabla\phi_2
\eea
 With this, the quadratic part of the action
 is given by:
\bea
{\mathcal S}_{0} + {\mathcal S}_{12} & = & \frac{J_s}{2 T} \int d^2r
 (\nabla \phi_s)^2 + \frac{J_a}{2 T} \int d^2r
(\nabla \phi_a)^2 \, ,
\label{Squadr}
\eea 
 where $J_s$ and $J_a$ are 
 given by $J_{s/a}=J \pm J_{{\rm int}}$.

 We now motivate the existence of such a term ${\mathcal S}_{12}$ in
 ultracold atom systems, by considering two BECs coupled
 by a short-range density-density interaction.
 Starting from a Hamiltonian of the 
 form $H = \sum_k [\epsilon_{\mathbf k} b^\dagger_{\mathbf k} b_{\mathbf k} 
 + (g/2V) \rho_{\mathbf k}^\dagger \rho_{\mathbf k}$],
 where $b_{\mathbf k}$ is the boson operator, $\epsilon_{\mathbf k}$ the free 
 dispersion $\epsilon_{\mathbf k}={\mathbf k}^2/2m$, $g$ is the interaction
  strength of the contact interaction, $V$ the volume,
 and $\rho_{\mathbf k}$ is the density operator of momentum 
 ${\mathbf k}$,
 given by $\rho_{\mathbf k}=\sum_{\mathbf p} b^\dagger_{\mathbf p} b_{\mathbf p+\mathbf k}$,
 we assume that the zero momentum  mode is macroscopically
occupied, and 
 formally replace the 
 operator $b_0$ by a number, 
 $b_0 \rightarrow \sqrt{N_0}$, where
 $N_0$ is the number of condensed atoms which is 
 comparable to the total atom number $N$, i.e. $N_0 \sim N$.
 Next we keep all terms that
 are quadratic in $b_\bold{k}$ (with $\bold{k}\neq0$), and
 perform a Bogoliubov transformation, given by:
 $b_\bold{k} = u_\bold{k} \beta_\bold{k} + v_\bold{k} \beta^\dagger_\bold{-k},$
 to diagonalize the Hamiltonian.
 The 
 eigenmodes $\beta_\bold{k}$ have a
 dispersion relation 
  $\omega_{\mathbf k} = \sqrt{\epsilon_{\mathbf k}(\epsilon_{\mathbf k} +2 g n)}$,
 with $n$ being the density $N/V$.
  The low-${\bf k}$ limit is given by
 $\omega_\bold{k}^2\sim v^2|{\bf k}|^2$, 
 with $v=\sqrt{g n/m}$, which 
 corresponds to the contribution in Eq. (\ref{S0}) of the action.
  Next, we consider 
 the sum of two copies of the previous Hamiltonian with boson operators
 $b_{1/2}$. 
 In addition we consider an interaction 
  $H_{12}= g_{12}/V \sum_{\mathbf k} \rho_{1, \mathbf k}^\dagger \rho_{2, \mathbf k}$,
 where the density operators 
 $\rho_{1/2, {\mathbf k}}$ are 
given by $\rho_{1/2, \mathbf k}=\sum_{\mathbf p} b^\dagger_{1/2, \mathbf p} b_{1/2, \mathbf p+\mathbf k}$.
 Following the same procedure as before, we find two eigenmode
 branches, corresponding to 
 in-phase and out-of-phase superpositions of the modes of
 each condensate, with
 the 
 dispersions $\omega_{s/a, \bold{k}}^2\sim v_{s/a}^2|{\bf k}|^2$,
 with the velocities  $v_{s/a}=\sqrt{(g \pm g_{12}) n/m}$.
 Therefore, for this example, the energy scale $J_{int}$ is related to
 $g_{12} n/m$, 
  which would be of similar order as $J$ for a system
 interacting via contact interaction, for small temperatures.
 This discussion only applies to the weakly interacting limit of a true 
condensate. However, it demonstrate that 
 a density-density contact interaction term can lead to a substantial 
 energy splitting of the
 in-phase and out-of phase modes. 

\begin{figure}[t]
\centerline{\includegraphics[width=6.0cm]{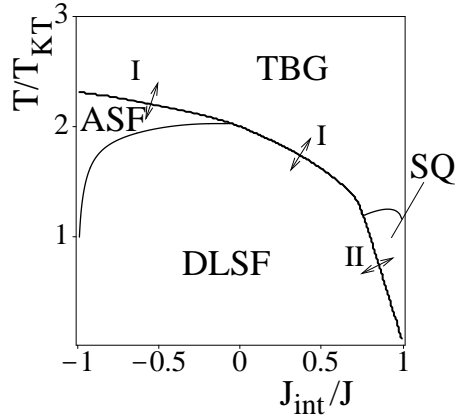}}
\caption{\label{pdint} Phase diagram, temperature (in units of
$T_{KT}$) versus interaction (in units of $J$). We assume $J_\perp
/J\sim 10^{-3}$ and $A_1/J \sim 10^{-3}$. DLSF: double layer
superfluid; TBG: thermal Bose gas; ASF: anti-symmetric superfluid;
SQ: symmetric quasi-order. The order of the transition
lines are either first (I), second order (II), or KT (thin lines).}
\end{figure}



Finally, in addition
 to single 
 vortices in each SF,
we have to consider the possibility of correlated vortex pairs, 
 i.e. one vortex in each layer at the same location of either
 the same 
 or of opposite vorticity.  
 We will refer to these vortex configurations as 
 {\it symmetric} or {\it anti-symmetric vortex pairs}, respectively. 
 These 
 excitations appear as the following terms
 in the action:
%
%
%
\bea\label{Srs}
{\mathcal S}_{s,a} & = & \frac{2 A_{s,a}}{T} \int \frac{d^2 r}{(2
\pi\alpha)^2} \cos(2\sqrt{2}\,\theta_{s,a}).
\eea
These 
 correlated vortex terms, 
 which describe new
degrees of freedom,
 can be the most relevant non-linear
terms in the action, 
%
 which derives from the
possibility that the vortices in different layers interact with each
other, through the
 terms (\ref{Sperp}) and (\ref{S12}). 
  The effect of these terms is the following:
 At low temperatures the energy
between two single vortices of opposite
 vorticity due to tunneling 
 grows as the
square of the distance D between them,
 i.e. as $J_\perp (D/\alpha)^2$. 
 As a result, the tunneling term attempts
 to confine vortices of opposite vorticity, leading
 to phase-locking between the layers, which we describe further
 later on.
%
 The interaction ${\mathcal S}_{12}$ changes
 the energy of correlated vortex pairs as follows: 
 The energy of a single vortex is given by
  $2\pi J\log L/\alpha$, where $L$ is the system size,
  whereas
  a symmetric/anti-symmetric vortex pair has
 an energy of 
 $4\pi(J\pm J_{int})\log L/\alpha$.
  Therefore, symmetric vortex pairs are the lowest energy vortex
 excitations for $J_{int}<-J/2$, whereas for $J_{int}>J/2$ anti-symmetric
 vortex pairs are the lowest energy excitations.
 As we will see below, in these regimes correlated vortex pairs 
 drive transitions to phases, in which one sector
 is (quasi-)SF whereas the other is disordered. 
 We  will also see that
 these terms are
generated under the RG flow, 
 even if not present at the onset.

 We note that a similar system has been studied in \cite{cazalilla}. Here,
 we consider a larger class of systems by including
 the interaction term (\ref{S12}), which in turn requires us to include
 the correlated vortex excitations (\ref{Srs}). These terms give rise
 to additional phases as we will see in the following.


Next we analyze our system within the RG approach. 
 This RG flow is perturbative in the vortex fugacities 
 $A_1$, $A_{s}$, and $A_a$, and the tunneling energy $J_\perp$,
 and therefore applies to the weak-coupling limit 
 (in particular 
 $J_\perp\rightarrow +0$).
 At second order
 the flow equations are given
by \cite{benfatto06}:
\bea
\label{RG}
&&\frac{d J_\perp}{d l}=\left(2-\frac{T}{2\pi J_a}\right) J_\perp \, ,
\label{JperpEqn}\\
&&\frac{d A_s}{d l}= \left(2-2\pi \frac{J_s}{T}\right) A_s +
\alpha_3\frac{A_1^2 (J_a -J_s)}{2 T^2} \, ,
\\
&&\frac{d A_a}{d l}= \left(2-2\pi \frac{J_a}{T}\right) A_a +
\alpha_3\frac{A_1^2 (J_s -J_a)}{2 T^2} \, ,
\\
&&\frac{d A_1}{d l}=\left(2-\frac{ \pi(J_s+ J_a)}{2T} + \alpha_3\frac{A_s J_s
    + A_a J_a}{T^2}\right) A_1 \, ,
\label{A1Eqn}
\\
&&\frac{d J_a}{d l}=\alpha_2 \Big(\frac{J_\perp^2}{4\pi^4 J_a} -
4\frac{A_a^2}{ T^4}  J_a^3 - \frac{A_1^2}{2 T^4}
(J_s+J_a)J_a^2\Big), \phantom{XX}
\\
&&\frac{d J_s}{d l}= - \alpha_2 \Big(2\frac{A_s^2}{ T^4}  J_s^2 +
\frac{A_1^2}{4 T^4} (J_s+J_a)J_s\Big) 2 J_s \, .
\eea
The coefficients $\alpha_{2/3}$ are non-universal parameters that
appear in the RG procedure~\cite{kogut79}, and which do not  affect the
results qualitatively. For consistency, we have to expand the
right-hand site of the above equations up to second order, around
the resulting Gaussian fixed point: $J_{s/a} = J \pm J_{{\rm int}} +
j_{s/a}$. We emphasize again that $J_{int}$ near the fixed point can
be generated by RG and be nonzero even if it is not present at the
onset.

Before we consider the full RG flow, we consider the 
 simpler case of no tunneling, i.e. 
 we solve the RG equations while setting 
$J_\perp =0$.
 In Fig. \ref{PD_Jint} we show the phase diagram of two 2D SFs coupled
 by 
 ${\mathcal S}_{12}$, Eq. (\ref{S12}).
  Such a system would be realized
 by a 2D mixture of bosonic atoms in two different hyperfine
 states, interacting via some short-range potential. 
 The order parameters we consider are $O_s(x)=b_1(x) b_2(x)$ 
and $O_a(x)=b_1^\dagger(x) b_2(x)$. 
 To obtain the phase diagram we consider
 the correlation functions of each of these order parameters,
 which can either scale algebraically or exponentially.
 In Fig. \ref{PD_Jint} we refer to algebraic scaling
 of $O_s(x)$ 
as symmetric quasi-order (SQ), 
 and 
 of $O_a(x)$
 as anti-symmetric quasi-order (AQ). 
 In each of the sectors a KT transition marks the 
 transition from the algebraic to the exponential regime,  
  which occur either simultaneously and are driven by single-vortex 
 excitations, or at different temperatures and are driven
 by correlated vortex pairs.
  As a result we find four regimes:
 At temperatures above $T_{KT}$, both sectors are disordered, giving
 rise to a thermal Bose gas (TBG) phase. For temperatures
 below $T_{KT}$, and for a wide range of $J_{int}$, we find
 that both sectors 
 are quasi SF
 (AQ/SQ), which
 is the only phase in which the correlation function of the
 single boson operators show algebraic scaling.
 We also find regimes in which only one sector shows
 algebraic scaling, whereas the other is disordered (AQ and SQ).
  From the perspective of vortices, the
 TBG phase is a gas of free single vortices in each layer,
 whereas the AQ (SQ) phase is a gas of symmetric (anti-symmetric) vortex pairs.

 We now consider the full RG system, including $J_\perp$. 
We numerically integrate the RG equations, and find the phase
diagram 
 shown in Fig. \ref{pdint} in terms of the
temperature $T$ and the interaction $J_{\rm int}$. We again
 find four
different phases that are different combinations of LRO,
 QLRO, and disorder in the symmetric and
anti-symmetric sector. 
 At high temperatures we find that both sectors are disordered
in a TBG phase, as before. 
 For lower temperatures, and 
 for a wide range of $J_{{\rm int}}$, 
the system is in a double-layer SF phase (DLSF): 
 The symmetric sector shows algebraic scaling, 
 whereas the exponent of the anti-symmetric sector is renormalized
 to zero, 
 i.e. we find two
SFs that are phase-locked due to 
 $J_\perp$. 
%
 Note that the transition
temperature $T_c$ between DLSF and TBG has been noticeably increased
relative to the decoupled value $T_{KT}$, as we will discuss further
later on. We also find two additional phases, which
 are partially (quasi-)SF and partially disordered. 
 One of them is the SQ phase, as before, whereas the other one (ASF),
 now shows true LRO in the anti-symmetric sector due to 
 $J_\perp$, whereas the
symmetric sector remains disordered. 
 We note that the generic double-layer
action that we discuss in this paper does not show a sliding
phase~\cite{toner}, for any non-zero $J_\perp$. Either 
$S_1$ or $S_a$, which is generated by RG, drives the anti-symmetric
sector to a disordered state, or $S_\perp$ creates true LRO in the
field $\phi_a$.

We also use the RG flow to find the order of the phase
transitions in the weak-coupling limit 
 that the anti-symmetric sector undergoes, by determining
the energy gap using a 'poor-man's scaling' argument: when the
coupling amplitude $J_\perp (l^*)$ is of order unity the
corresponding gap is given by the expression $\Delta \sim J_\perp
\exp(-l^*)$. From the behavior of 
 $\Delta$
 at the phase transition
we can read off whether it is of first or second order, as indicated
in Fig.~\ref{pdint}.
\begin{figure}
\centerline{\includegraphics[width=8.0cm]{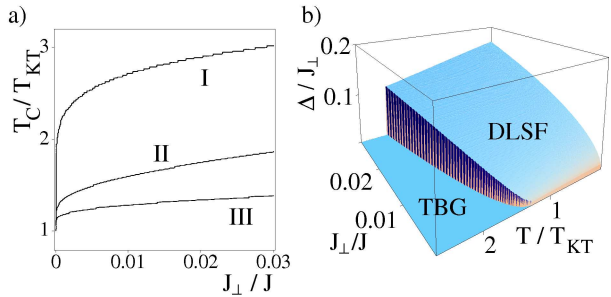}}
\caption{\label{PD} (a) Critical temperature $T_c$ of the 
 DLSF-TBG transition (in units of $T_{KT}$)
for different values of $A_1/J$: $10^{-3}, 0.1, 0.4$ (I--III), and
 for $J_{\rm int} =0$. (b)
Energy gap in the anti-symmetric sector (in units of $J_{\perp}$) as
a function of $J_{\perp}/J$ and temperature (in units of $T_{KT}$).
We have set $A_1/J=0.1$
 and $J_{int}=0$.}
\end{figure}

Given the nature 
 of an effective theory, only
approximate statements can be made about how the different regimes
of the phase diagram relate to the microscopic interactions. To
create the ASF or AQ phase an attraction between the two
  atom species is needed 
 that is of order $J$, whereas to create the SQ
phase, a repulsion of that order would be needed.
 To detect the different phases, one could use the interference method
 used in \cite{zoran} to distinguish the phase-locked
 phases (DLSF and ASF), which would show a well-defined interference
 pattern, from the uncorrelated phases. Another approach would be
 time-of-flight images: The DLSF phase would display a quasi-condensate
 signature, whereas the other phases would appear disordered. However,
 at the transition from ASF 
 or SQ to TBG, the width of the
 distribution would abruptly increase.

\subsection*{\textcolor{NavyBlue}{Kibble-Zurek mechanism}}
\addcontentsline{toc}{subsection}{Kibble-Zurek mechanism}

In this section we discuss how the phase-locking transition
 found in the previous section could be used to realize the KZ mechanism.
  The defining property of this mechanism 
 is the generation of topological defects
 by ramping across a phase transition, coming from the disordered
 phase. The disordered phase that
 we propose to use 
is the 
  TBG phase
 of the {\it decoupled} 2D systems, that is, we consider
  the experimental setup  reported  in ref.~\cite{zoran} for
 a temperature $T$ above the KT temperature $T_{KT}$.
  The ordered phase we consider is the DLSF phase, i.e. 
  the phase-locked phase of two {\it coupled} SFs. 
 The ramping is achieved by turning on the tunneling between the two layers,
 which can be done by lowering the potential barrier
 between them. 
  For this procedure the critical 
 temperature $T_c$ of the DLSF-TBG transition needs to be
 above the KT temperature of the uncoupled systems.
  We now show that the RG flow indeed predicts such a scenario. 
 In the experiments in Ref.~\cite{zoran}, the atoms
in different layers do not interact with each other. Therefore, it
can be expected that $J_{\rm int}$ is small,  of order 
 $J_\perp$, 
 which motivates
 us to discuss the case $J_{int}=0$ here. 
 We note however, that the desired scenario of an increased
 critical temperature,  
 is found for a wide range of $J_{int}$, as
 can be seen in Fig. \ref{pdint}. 
 In Fig. \ref{PD} (a) we show how the critical temperature
of the DLSF-TBG transition behaves, predicted by the RG flow, for
different values of $A_1$.
 The critical temperature shows a sizeable
increase, due to the phase-locking transition. Due to the
perturbative nature of the RG scheme, the RG flow underestimates the
effects of the term $S_1$, and predicts a finite jump of the
critical temperature when $J_\perp$ is turned on. However, to lock
the SFs together in the regime slightly above $T_{KT}$,
$J_\perp$ needs to be at least of the order of the vortex core
energy, giving rise to a finite slope of $T_c$ instead of a jump.
The energy gap of this transition is shown in Fig.~\ref{PD} from
which we can see that the transition is of first order, 
 in contrast to the second order transition 
 described in \cite{kibble, zurek}, which
 is advantageous because the onset of order is instantaneous
 rather than continuous. 
 We
note that the phase diagram was obtained using the assumption that
the bare parameters of the model, in particular 
 $J$, do not depend on temperature. This is true only if
temperature is close to $T_{KT}$. Here we find that the ratio
$T_c/T_{KT}$ can be relatively large. In fact $T_c/T_{KT}$ will be
always smaller than that shown in Fig~\ref{PD} (a), however,
qualitatively the behavior of $T_c/T_{KT}$ as a function of
$J_{\perp}$ should remain intact. We point out that our results can be
generalized to a system of $N>2$ coupled SFs. One finds that the SFs
still show a strong tendency to phase-lock together. As a result
the critical temperature should approximately satisfy the equality
$\pi J(T_c) N =2 T_c$. Thus as $N$ increases $T_c$ approaches the
mean-field critical temperature at which the stiffness $J$ vanishes
and we recover the usual 3D result.

In finite size systems there is another constraint on the minimum
value of $J_{\perp}$:
 We consider the free energy of a single vortex in the anti-symmetric
field: $\phi_a \sim \arctan(x/y)$. For the decoupled system we get
 for the free energy \cite{kogut79}: 
 $F \sim 2(\pi J -2 T)\log L/\alpha$, where
 $L$ is the system size. 
 The coupling term
gives a free energy 
 contribution $F_{\perp}\sim J_\perp (L/\alpha)^2$. In the
thermodynamic limit, $L \rightarrow \infty$, this term diverges
faster than the others, which is consistent with our finding of LRO
in the antisymmetric sector. For a finite system,  comparing 
these terms gives the estimate $J_\perp\sim J
\log(L/\alpha)/(L/\alpha)^2$, that is required for this order to
develop.
 With a system size $L/\alpha \sim 10^2$, that 
 would require $J_\perp\sim 10^{-3} J$, which, for the 
 setup in \cite{zoran}, would be around $10^2 s^{-1}$.

As an estimate of the number  of domains that would be created, we
 follow the argument in \cite{kibble}:
 The coherence scale of the DLSF phase is given by $(J/\Delta)^{1/2}\alpha$,
 which is the scale 
  of a Klein-Gordon model with
 a kinetic energy scale $J$ and a 'mass-term' with a prefactor
 $\Delta/\alpha^2$. 
 The domain size is then given by $(J/\Delta)\alpha^2$, and the
 number of domains 
 by $\sim (\Delta/J) L^2/\alpha^2$.
  As we show in Fig. \ref{PD} b) for $J_\perp/J\approx 10^{-2}$, 
 we find $\Delta/J_\perp \sim 10^{-1}$, and therefore 
 $J/\Delta \sim 10^3$. With $L/\alpha \sim 10^2$,
  we would get $N_{dom} \sim 10^1 - 10^2$, which would generate a 
 similar number of vortices. 
  We estimate the vortex-antivortex imbalance by considering
 the number of domains around the periphery of the 
 system, which scales as $L/\xi$. If we imagine that the
 phase behaves like a random walk, the total phase mismatch, corresponding
 to the vortex-antivortex imbalance, 
 will scale as $\sqrt{L/\xi} \sim N_{dom}^{1/4}$, 
 which, for $L/\alpha\sim 10^2$,
 is of the order $10^0 - 10^1$.

In summary, we propose the following procedure: 
 i) Prepare two uncoupled SFs at a temperature $T$ slightly above 
 $T_{KT}$.
  ii) Switch on the tunneling between the two layers, which creates a DLSF 
 phase with a critical temperature $T_c$ higher than $T$.
 As a result, one should find a number of long-lived 
 vortex-antivortex pairs in the 
 anti-symmetric phase field $\phi_a$, which would 
 be visible in an interference measurement, at
 a temperature where there would be none in thermal equilibrium.

In conclusion of the section,
  we studied the phase-locking transition of 2D
superfluids, within an renormalization group approach. 
 We find that this transition is accompanied by an increase of the transition
temperature. We suggest that this effect can be used to probe the
Kibble-Zurek mechanism in cold atom systems by rapidly changing the
ratio $T/T_c$. When we include interactions between the layers  
 we find  additional phases, in which  either the symmetric
or the anti-symmetric sector is disordered, and
 the other sector stays superfluid or quasi-superfluid.

\section*{\textcolor{NavyBlue}{V. Bose-Fermi mixtures
 in two-dimensional optical lattices}}
\addcontentsline{toc}{section}{V. Bose-Fermi mixtures
 in two-dimensional optical lattices}
%
%
In the spirit of engineering many-body systems 
 that are relevant in other fields, we now turn 
 to atomic mixtures that resemble qualitatively, i.e. in terms of degrees of freedom 
 of the system,  electron-phonon systems.
 In two dimensions, these systems are
 actively studied and prove to be of considerable complexity. 
 In order to study their atomic counterparts, Bose-Fermi mixtures in 
 optical lattices, we use the
 powerful method of functional renormalization group equations, 
 with which we can determine their phase diagrams
 in the weak-coupling limit in a systematic fashion.
 We find a rich competition of phases for both the square lattice
 and triangular lattice geometry that we consider. 

In this section we consider 
  mixtures of one bosonic type of atom and either 
 two fermionic types that are SU(2) symmetric or
 spinless fermions. The Hamiltonian 
 for a mixture on a square lattice is given by:
\bea\label{Ham}
H \!= \!-t_{f}  \sum_{\langle i j \rangle,s} f^\dagger_{i,s} 
f_{j,s} \!-\! t_{b}  \sum_{\langle i j \rangle} b^\dagger_{i} b_{j}
\!-\! \sum_i (\mu_{f} n_{f,i}\! +\! \mu_b n_{b,i})\!+\! \sum_{i} \Big[U_{ff} n_{f, i,\uparrow}n_{f,i, \downarrow}
\!+\! \frac{U_{bb}}{2} n_{b, i}n_{b,i} \!+\! U_{bf} n_{b, i} n_{f,i} \Big] \, ,
\eea
where $f^\dagger_{i,s}$ ($f_{i,s}$) creates (annihilates) a fermion at site $i$ with pseudo-spin $s$ ($s=\uparrow,\downarrow$), $b^\dagger_i$ ($b_i$) creates (annihilates) a boson at site $i$,
$n_{f,i}= \sum_s f^\dagger_{i,s} f_{i,s}$ ($n_{b,i} = b^\dagger_i b_i$) is the fermion (boson) number operator,  $t_f$ and $t_b$ are the fermionic and bosonic tunneling energies between neighboring sites, 
$\mu_f$ ($\mu_b$) is the chemical potential for fermions (bosons), 
 $U_{bb}$ is the repulsion energy between bosons on the same site, $U_{ff}$ is the repulsion energy between the two species of fermions, and $U_{bf}$ is the repulsion
energy between bosons and fermions. The two fermion species have been treated as a pseudo-spin-$1/2$ index ($\uparrow$ and $\downarrow$). The case of spinless fermions can be immediately obtained from (\ref{Ham}) by ignoring one of the spin states.
In momentum space, the Hamiltonian (\ref{Ham}) is written as:
\bea\label{Ham_k}
H\! =\! \sum_{\mathbf{k}} \left\{ \!(\epsilon_{f, \mathbf{k}} \!-\!\mu_f) \sum_s f^\dagger_{\mathbf{k},s} f_{\mathbf{k},s} \!+\! (\epsilon_{b, \mathbf{k}}\! -\!\mu_b) 
b^\dagger_{\mathbf{k}} b_{\mathbf{k}}\! +\!
\frac{U_{ff}}{V} \rho_{f,\mathbf{k},\uparrow}\rho_{f,-\mathbf{k}, \downarrow}
 \!+\! \frac{U_{bb}}{2 V}  \rho_{b, \mathbf{k}}\rho_{b,-\mathbf{k}}  \!+\! \frac{U_{bf}}{V} \rho_{b, \mathbf{k}}
\rho_{f,-\mathbf{k}}\! \right\}  \, ,
\eea
where $\rho_{f,{\bf k}} = \sum_{{\bf q},s}  f^\dagger_{\mathbf{k}+{\bf q},s}
f_{\mathbf{k},s}$ ($\rho_{b,{\bf k}} = \sum_{{\bf q}}  b^\dagger_{\mathbf{k}+{\bf q}} b_{\mathbf{k}}$) is the fermion (boson) density operator,  $\epsilon_{b/f, \mathbf{k}}= - 2 t_{b/f} (\cos k_x + \cos k_y)$, is the bosonic/fermionic dispersion relation. 

 We consider the  
limit  of weakly interacting bosons that form a BEC \cite{wang05,mathey06a}, 
 where we assume that the zero momentum bosonic mode is macroscopically
occupied, and the corresponding operator
$b_0$ can be formally replaced by a real number $b_0 \rightarrow \sqrt{N_0} $, where
$N_0$ is the number of condensed atoms.
After this replacement we keep all terms that
are quadratic in $b_\mathbf{k}$ (with $\mathbf{k}\neq0$), and
perform a Bogoliubov transformation, given by:
$b_\mathbf{k} = u_\mathbf{k} \beta_\mathbf{k} + v_\mathbf{k} \beta^\dagger_\mathbf{-k},$
to diagonalize the bosonic Hamiltonian.
The resulting eigenmodes $\beta_\mathbf{k}$ have a
 dispersion relation given by 
 $\omega_\mathbf{k} = \sqrt{\epsilon_{b,\mathbf{k}}(\epsilon_{b,\mathbf{k}} +2U_{bb}n_b)}$ ,
with the low-${\bf k}$ limit $\omega_\mathbf{k}\sim v_b|{\bf k}|$, 
with $v_b=\sqrt{2 t_b U_{bb} n_b}$.
The parameters $u_\mathbf{k}$ and $v_\mathbf{k}$ are given by:
$u_\mathbf{k}^2  =  (\omega_\mathbf{k} +\epsilon_{b,\mathbf{k}} + U_{bb} n_b)/(2 \omega_\mathbf{k})$ and 
$v_\mathbf{k}^2  =  (-\omega_\mathbf{k} + \epsilon_{b,\mathbf{k}} + U_{bb}n_b)/(2\omega_\mathbf{k})$.

The density fluctuations of the bosons are approximated by:
$\rho_{b,\mathbf{k}}\approx \sqrt{N_0}(u_\mathbf{k} +v_\mathbf{k})
(\beta_\mathbf{k} + \beta_{-\mathbf{k}}^\dagger)$, with $\mathbf{k}\neq 0$.
 The interaction between bosons and fermions is then given by 
 $U_{bf}\sqrt{N_0}/V\sum_{\mathbf{k}}(u_\mathbf{k}+v_\mathbf{k})(\beta_\mathbf{k} 
+\beta_{-\mathbf{k}}^\dagger) \rho_{f,-\mathbf{k}}$.
As a next step we integrate out the bosonic  modes and use an instantaneous  approximation, leading to the following effective
Hamiltonian:
\bea\label{Heff}
H_{{\rm eff.}}  =  \sum_{\mathbf{k} } \left\{ (\epsilon_{\mathbf{k}} -\mu_f) 
\sum_s f^\dagger_{\mathbf{k},s} 
f_{\mathbf{k},s} + \frac{U_{ff}}{V} \rho_{f,\mathbf{k},\uparrow}\rho_{f,-\mathbf{k}, \downarrow} +  \frac{1}{2V} V_{{\rm ind.}, \mathbf{k}} \, \, \rho_{f,\mathbf{k},} \rho_{f,\mathbf{-k}} \right\}
\, ,
\eea
where the induced potential $V_{{\rm ind.}, \mathbf{k}}$ is given by:
\bea\label{vind}
V_{{\rm ind.}, \mathbf{k}}= - \tilde{V}/(1 + \xi^2 (4 - 2 \cos k_x -2 \cos k_y)) \, ,
\eea
with $\tilde{V}$ given by $\tilde{V}=U_{bf}^2/U_{bb}$, and $\xi$ is the
 healing length of the BEC and is given by  $\xi=\sqrt{t_b/2 n_b U_{bb}}$. This approach is
 only valid when $v_b\gg v_f$, so that the fermion-fermion interaction mediated by the bosons can be considered as instantaneous. Away from this limit, retardation effects are present. In this case, one has to consider the frequency dependence
 of the interaction explicitly \cite{klironomos07,tsai05a,tsai05b}. The full effective interaction, 
 including retardation, is given by:
 \bea
 V_{\rm ind.}(\omega,\mathbf k)= - \left[\frac{\tilde{V}}
{1+\xi^2(4-2\cos{k_x}-2\cos{k_y})}\right]
\frac{\omega^2_{\mathbf k}}{\omega^2
+\omega^2_{\mathbf k}},
 \eea
and the static limit (\ref{vind}) is recovered when $\omega_{\mathbf k} \gg \omega$.
Equation (\ref{Heff}) describes the scattering of two fermions from momenta ${\bf k}_1$ and ${\bf k}_2$, that are scattered into momenta 
${\bf k}_3$ and ${\bf k}_4$. Momentum conservation at the interaction vertex
requires that ${\bf k}_4 = {\bf k}_1+{\bf k}_2 -{\bf k}_3$, and hence the
interaction vertex, $U(\mathbf{k}_1,\mathbf{k}_2,\mathbf{k}_3)$, depends on
three momenta. Its bare value from (\ref{Heff}) can be written as:
\bea
U(\mathbf{k}_1,\mathbf{k}_2,\mathbf{k}_3) & = & U_{ff} + V_{{\rm ind.}, \mathbf{k}_1-\mathbf{k}_3} \, .
\label{Uk}
\eea
For the case with retardation, there is dependence on both the momenta and frequencies of the electrons so we have $U(k_1,k_2,k_3)$, with $k_4=k_1+k_2-k_3$, where $k_i=(\omega,{\mathbf k})$.

Starting from non-interacting fermions, we ask the general question of what new many-body phases can emerge when the system is subjected to a given interaction $U(\mathbf{k}_1,\mathbf{k}_2,\mathbf{k}_3)$. 
 Our approach to address this question is the renormalization-group method, described in the next section.

\subsection*{\textcolor{NavyBlue}{Renormalization-Group Method}}
\addcontentsline{toc}{subsection}{Renormalization-Group Method}

Starting with a microscopic model of interacting electrons on a lattice, the 
renormalization-group (RG) method provides the effective model at a given
temperature or energy scale \cite{shankar94}. 
The RG is implemented by systematically tracing out high energy degrees of freedom in a region between $\Lambda$ and $\Lambda + d\Lambda$, where 
$\Lambda$ is the energy cut-off of the problem. In this process, the vertex
$U$ is renormalized. At the initial value of the cut-off $\Lambda=\Lambda_0$, the value of $U$ is given by its bare value. For the BFM system we describe here it is given by (\ref{Uk}). At one loop, the RG flow is obtained from a series of coupled
integral-differential equations \cite{zanchi00} for all the different
interaction vertices $U(\mathbf{k}_1,\mathbf{k}_2,\mathbf{k}_3)$. 
The RG equations read:
\begin{eqnarray}
\partial_{\ell} U_{\ell}(\!{\bf k_1},\!{\bf k_2},\!{\bf k_3}\!) \!=\!
-\!\!\!\int_{p,\omega} \!\!\partial_{\ell}
[G_{p\ell}G_{k\ell}] U_{\ell}(\!{\bf k_1},\!{\bf k_2},\!{\bf k}\!)
U_{\ell}(\!{\bf p},\!{\bf k},\!{\bf k_3}\!)\!-\!\!\int_{p,\omega} \!\!\partial_{\ell}
 [G_{p\ell}G_{q_1\ell}]
U_{\ell}(\!{\bf p},\!{\bf k_2},\!{\bf q_1}\!) U_{\ell}(\!{\bf k_1},\!{\bf q_1},\!{\bf k_3}\!)&&
\nonumber\\
\!\!-\!\!\!\int_{p,\omega} \!\!\!\!\partial_{\ell}
 [G_{p\ell}G_{q_2\ell}] \!\!\left\{-\!2U_{\ell}(\!{\bf k_1},\!{\bf p},\! {\bf q_2}\!)U_{\ell}(\!{\bf q_2},\!{\bf k_2},\!{\bf k_3}\!) 
\!+ \! U_{\ell}(\!{\bf p},\!{\bf k_1},\!{\bf q_2}\!) U_{\ell}(\!{\bf q_2},\!{\bf k_2},\!{\bf k_3}\!)\! 
+\!  U_{\ell}(\!{\bf k_1},\!{\bf p},\!{\bf q_2}\!)
U_{\ell}(\!{\bf k_2},\!{\bf q_2},\!{\bf k_3}\!) \!\right\}&&
\end{eqnarray}
where $\ell = \ln(\Lambda_0/\Lambda)$, ${\bf k} = {\bf k}_1+{\bf k}_2-{\bf p}$, ${\bf q}_1={\bf p}+{\bf k}_2-{\bf k}_3$, ${\bf q_2}={\bf p}+{\bf k}_1-{\bf k}_3$, and 
$G_{k\ell} = \Theta(|\xi_{{\bf k}}|-\Lambda)/(i \omega-\xi_{{\bf
    k}})$ with $\xi_{\bf k} = \epsilon_{f,{\bf k}} - \mu_f$ and $k=(\omega,\bf k)$.

From the general interaction vertices $U(\mathbf{k}_1,\mathbf{k}_2,\mathbf{k}_3)$, the
specific interaction channels, such as charge-density wave (CDW), antiferromagnetic (AF),
and  superconducting (BCS), can be obtained:
\bea
\label{eq:cdw} V^{CDW} & = & 4 \, \, U_c(\mathbf{k}_1,\mathbf{k}_2, \mathbf{k}_1+\mathbf{Q}) \, ,
\\
\label{eq:af} V^{AF} & = & 4 \, \, U_\sigma(\mathbf{k}_1,\mathbf{k}_2, \mathbf{k}_1+\mathbf{Q}) \, ,
\\
\label{eq:bcs} V^{BCS} & = & U(\mathbf{k}_1,-\mathbf{k}_1, \mathbf{k}_2) \, ,
\eea
where we have used the notation: 
$U_c  =  (2-\hat{X})U/4$, $U_\sigma =  - \hat{X}U/4$ 
with $\hat{X}U(\mathbf{k}_1,\mathbf{k}_2,\mathbf{k}_3)=U(\mathbf{k}_2,\mathbf{k}_1,\mathbf{k}_3)$, and $\mathbf{Q}$ is the nesting vector, $\mathbf{Q}=(\pi,\pi)$.  

In a numerical implementation, one discretizes the Fermi surface into $M$ patches,
and hence each of the interaction channels (\ref{eq:cdw}), (\ref{eq:af}),
(\ref{eq:bcs}) is represented by an $M\times M$ matrix. 
At each RG step, we
diagonalize each of these matrices. The channel with the largest
eigenvalue (with the caveat that a BCS-channel needs to be attractive to
drive a transition) corresponds to the dominant order. The elements of the
eigenvector are labeled by the discrete 
patch indices around the Fermi surface and the symmetry of the order
parameter is given by this angular dependence. 

The RG method for interacting fermions has been extended to also include retardation effects, as for the case of interacting electrons which are also coupled to phonons in a crystal \cite{tsai05a,tsai05b}. In this case: (i) the interaction vertices also depend on frequencies of the incoming and outgoing fermions, so the RG equations are written for given external frequencies and the integral over intermediate frequencies can not be done analytically, and (ii) there are important self-energy corrections (in particular the imaginary part of the self-energy is non-zero). Eliashberg equations for strong-coupling superconductivity has been derived with this method \cite{tsai05a,tsai05b} for the case of electrons, with a circular Fermi surface, coupled to phonons. This method has also been applied to other electron-phonon problems \cite{tam07,klironomos06}, and to mixtures of cold atoms in an optical lattice \cite{klironomos07}.

\subsection*{\textcolor{NavyBlue}{Phase Diagram and Sub-Dominant Orders}}
\addcontentsline{toc}{subsection}{Phase Diagram and Sub-Dominant Orders}

\begin{figure}[h]
\centerline{\includegraphics[width=7.5cm]{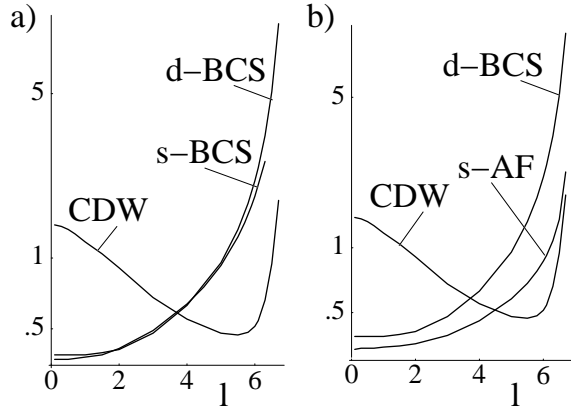}}
\caption{\label{Flow_HF}
RG flow for the different effective interactions (in units of
$t_f$) as a function of the RG parameter $l$ ($\tilde{V}/t_f=3$ and $\xi=1$). a) $U_{ff}/t_F =0.5$; b) $U_{ff}/t_f=1.2$.}
\end{figure}

The microscopic parameters in the Hamiltonian (\ref{Ham}) determine the initial conditions for the RG flow, and the shape of the Fermi surface. With these, we write the RG flow equations and solve them numerically. We first discuss the case without retardation. For some parameters, we encounter a divergence in the RG flow, indicating the onset of ordering with a gap that is in the detectable regime, i.e. larger than $10^{-3} t_f$. In other regimes, where such a divergence is not reached,  one can read off the dominant tendency of the RG flow.
In Fig.~\ref{Flow_HF} we show examples of RG flows as a function of
$\ell$. In Fig.~\ref{Flow_HF} (a), we show the competition between d-wave and
s-wave pairing, with d-wave being dominant and s-wave being
 subdominant. In Fig.~\ref{Flow_HF} (b) we show an example with dominant d-wave channel and subdominant  AF channel. In both cases we find that  for short distances (or high energies) CDW fluctuations are dominant, giving rise to a state that resembles the findings for high-$T_c$ superconductors \cite{checkerboard1,checkerboard2,checkerboard3}. In some situations the many-body states are almost degenerate and small changes in the initial conditions (that is, changes in the form of the interactions) can be used to select one particular ground state. 

\begin{figure}[h]
\centerline{\includegraphics[width=8cm]{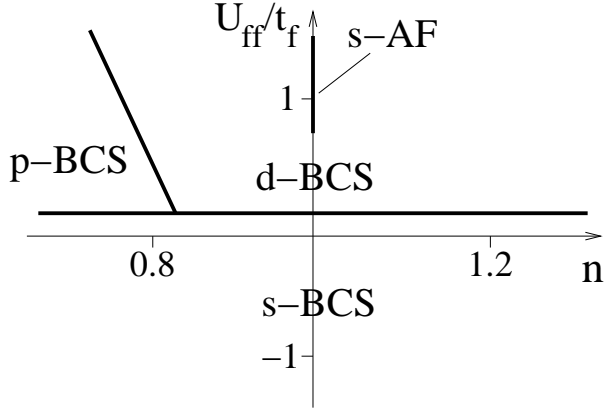}}
\caption{\label{PDsquare}
Phase diagram, interaction strength, $U_{ff}/t_f$, versus number of fermions per site, $n$, for a Fermi-Bose mixture in a square lattice
in 2D ($\tilde{V}/t_f = 2$, $\xi = 1$).}
\end{figure}

With this procedure we determine
 the phase diagram of the system, which is shown in Fig.~\ref{PDsquare}.
We  now discuss the general features of the phase diagram.
In the absence of any coupling to the bosons, i.e. for $\tilde{V}=0$, the
system shows s-wave pairing for attractive interaction, $U_{ff}<0$,
and no ordering for $U_{ff}>0$, i.e. Fermi liquid behavior, except for the special case of half-filling where Fermi surface nesting 
drives the system to AF order 
for repulsive 
interactions, and to s-wave pairing (degenerate with CDW) 
for attractive interaction. 
If we now turn on the interaction to the bosons, this picture is modified in
 the following way: The boundary of the s-wave regime is moved into the
 regime of positive $U_{ff}$, approximately to a value of $U_{ff}$ where the
 effective interaction at the nesting vector ${\bf Q}$ between the fermions,
 $U_{ff}+V_{{\rm ind.},\mathbf{Q}}$, is positive, i.e. for $U_{ff}\approx \tilde{V}/(1+8\xi^2)$. On the repulsive side, and
 away from half-filling,  we find the tendency to form a paired state, either
 $d$-wave or $p$-wave. This tendency becomes weaker the further the system is
 away from half-filling.  We typically find a gap in the vicinity of
 half-filling and further away from $\mu=0$ we find only an increasing strength of the corresponding interaction channel.  For the half-filled system, we find that for attractive interactions the degeneracy between $s$-wave pairing and CDW ordering is lifted, with $s$-wave pairing being the remaining type of order. For repulsive interactions, we find an intermediate regime of $d$-wave pairing, and for larger values of $U_{ff}$ we obtain AF order. 

\begin{figure}[h]
\centerline{\includegraphics[width=8.5cm]{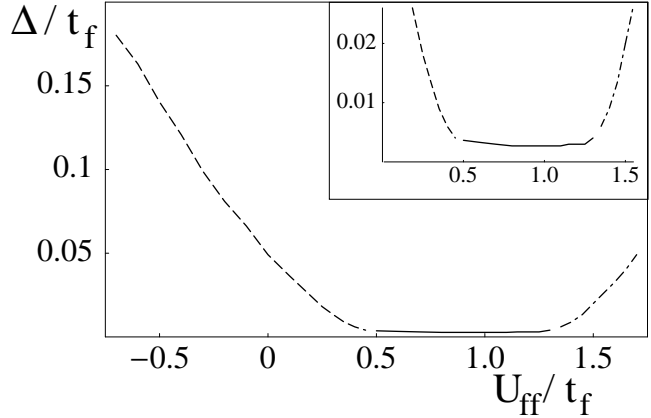}}
\caption{\label{Gap_HF}
Many-body energy gaps at half-filling ($\mu=0$), as a function
 of $U_{ff}/t_f$, for $\tilde{V}/t_f=3$ and at a fixed value
 of the coherence length $\xi=1$. Dashed line: s-wave gap; Continuous line: d-wave gap; Dotted-dashed line: antiferromagnetic gap.  The inset shows a magnified plot of the d-wave regime.}
\end{figure}

The RG approach also allows the extraction of the many-body gaps
in the system through a "poor man's scaling" analysis of the divergent flow: at the point where the coupling becomes of order of $t_f$
the scaling parameter $\ell$ reaches the maximum value $\ell_c 
= \ln(t_f/\Delta)$, where $\Delta$ is the value of the gap. Hence,
$\Delta/t_f \approx \exp\{-\ell_c\}$ can be obtained from the RG
flows such as the ones in Fig.~\ref{Flow_HF}. In Fig.~\ref{Gap_HF} we show the gaps of the problem as a function of $U_{ff}/t_f$ in the half-filled case. One can see that as $U_{ff}$ increases, from negative to positive values, the $s$-wave gap is replaced by a $d$-wave gap, and finally for an antiferromagnetic gap. As is apparent from this figure, the gap in the $d$-wave phase is much smaller than the gaps of the AF order and the $s$-wave pairing, and, furthermore, almost independent
 of the value of $U_{ff}$. The latter is the case because the $U_{ff}$ term is a pure $s$-wave contribution to the interaction and therefore does not contribute to the $d$-wave  channel. The $d$-wave channel has an initial contribution which is entirely due to the anisotropy of the induced interaction, which gives only a small   value, and as a consequence only a small value for the gap. The value of the gap (in units of $t_f$) can be numerically fitted with a
 BCS expression of the form $a \exp(-b/\tilde{V})$, with the parameters $a$ and $b$ given by $a=0.31$ and $b=14.2$.

For a system of spinless fermions, one can simply suppress one of the spin indices in (\ref{Ham}) and (\ref{Ham_k}). In this case there is a major simplification in the problem since $U_{ff}$ is absent: in a spinless problem there can be only one fermion per site, as per Pauli's principle. Hence, in the absence of bosons, the spinless gas is non-interacting. The bosons, however, mediate the interaction between the fermions. Since the fermions are in different lattice sites the pair wavefunction has necessarily a node and hence, no $s$-wave pairing is allowed. In other words, in the spinless case the anti-symmetry of the wavefunction requires pairing in an odd angular momentum channel. In fact, we find that throughout the entire phase diagram the fermions develop $p$-wave pairing. At half-filling we find
a similar behavior of CDW fluctuations on short scales, analogous to the flow shown in Fig. \ref{Flow_HF}.  One should point out that in real solids the conditions of "spinlessness" behavior is hard to achieve since it usually requires complete polarization of the electron gas, that is, magnetic energies of the order of the Fermi energy (a situation experimentally difficult to achieve in good metals). However, in cold atom lattices this situation can be easily accomplished with the correct choice of atoms.

\begin{figure}[h]
\centerline{\includegraphics[width=7.5cm]{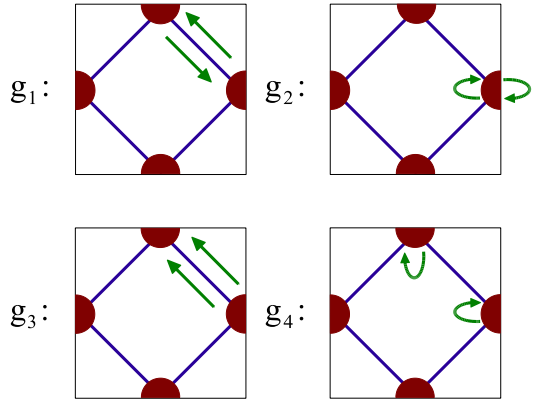}}
\caption{\label{vanhove}
Relevant processes in the two-patch approximation.}
\end{figure}

\begin{figure}[h]
\centerline{\includegraphics[angle=-90,totalheight=5.5cm,width=7.0cm,viewport=5 5 560
690,clip]{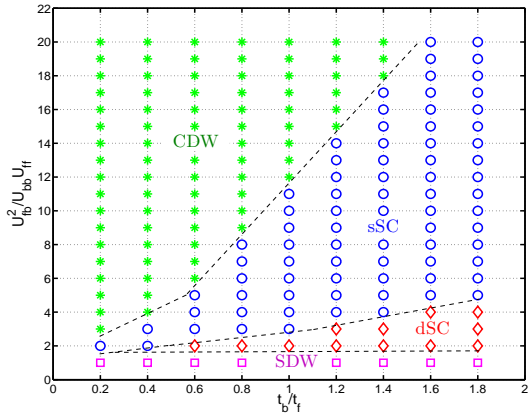}}
\caption{(Color online) Phase diagram for $U_{ff}=0.4t_f$, $U_{bb}=0.8t_f$,
and $n_b=2.5$.
Blue circles indicate $s$-wave SC, red rhombuses indicate $d$-wave SC, magenta squares AF (also called spin-density-wave SDW), 
and green stars CDW type of ordering. Dashed lines are guides to the eye.}
\label{PDretarded}
\end{figure}

Finally, when retardation is important, the numerical task of solving the RG flow equations become much more demanding. In addition to the discretization of the Fermi surface, one has to also discretize the frequency (for $T=0$), or consider a certain number of Matsubara frequencies ($T\ne 0$). This has been done for this Bose-Fermi system only for a fixed density of fermions corresponding to one half \cite{klironomos07}. In this case the Fermi surface has a diamond shape and scattering processes are dominated by the van Hove points (corners of the diamond) where the density of states is singular. The Fermi surface can therefore be approximated by the van Hove points only \cite{schulz87} so that he types of relevant processes are reduced, as shown in Fig. \ref{vanhove}. Each of the processes still depend on frequencies: $g_i(\omega_1,\omega_2,\omega_3)$, for $i=1,2,3,4$. The phase diagram is shown in Fig. \ref{PDretarded}. Retardation leads to additional phases at half-filling and by tuning the lattice parameters, the system undergoes AF (or spin-density-wave SDW), $d$-wave SC pairing, $s$-wave-pairing, and CDW. In the limit of $v_b \gg v_f$ discussed previously, when retardation is not important, CDW does not become dominant (Fig. \ref{PDsquare}). It is at most degenerate with $s$-wave pairing for $U_{ff} < 0$. As the bosons become slower, there is stronger tendency for CDW formation (Fig. \ref{PDretarded}).

\subsection*{\textcolor{NavyBlue}{Quantum Frustration in Triangular Lattices}}
\addcontentsline{toc}{subsection}{Quantum Frustration in Triangular Lattices}

It is known that the geometric shape of the lattice is a crucial factor
in determining the properties of interacting many-body systems. For instance,
localized spins interacting antiferromagnetically on a triangular lattice
suffer from the phenomenon of {\it frustration}, when antiferromagnetic order cannot
be achieved because of the particular lattice structure. 
For itinerant fermionic systems, the lattice structure, together with the 
dispersion relation and the filling fraction, determine the shape of the
Fermi surface. 
The Fermi surface, by its turn, is a crucial factor in determining what type of orders the
system can develop. Indeed, for the triangular lattice we consider in this
Section, which shows a rich and subtle competition between superconducting phases
with different symmetries, small changes in the shape of the FS determine which pairing symmetry
is dominant. This is a reflection of the ``lattice frustration'' on the
superconducting phases.
In solids, this intriguing lattice geometry is realized in materials such as cobaltates
\cite{cobaltates}, transition metal dichalcogenides \cite{dichal} and $\kappa$-(ET)$_2$X
layered organic crystals  \cite{jerome} (if each lattice site is represented by one ET dimer
\cite{kino}), and has been the subject of several theoretical studies \cite{tsai01,honerkamp03,vojta99,baskaran03,lee05,choy05,gan06,zhou06}

\begin{figure}[h]
\centerline{\includegraphics[width=8.0cm]{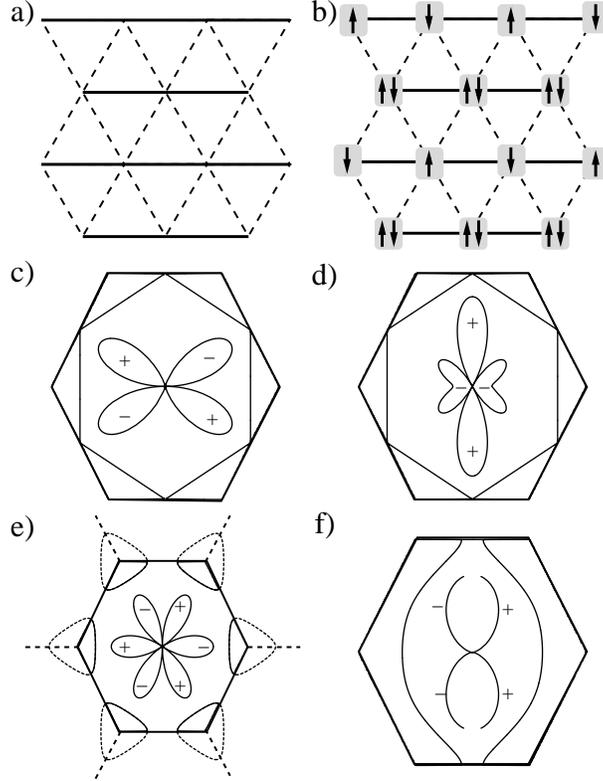}}
\caption{\label{phases}
 a) Lattice geometry of the system.  
 The continuous (dashed) bonds correspond to the hopping amplitudes
 $t_{b/f, 1 (2)}$. 
 For $t_{b/f, 1}=t_{b/f, 2}$, the lattice is an isotropic triangular lattice.
  b) schematic representation of the AF order corresponding to nesting
 vector ${\bf Q}_1$.  
 c) + d) Order parameters of  the extended $d$-wave orders $D_1$ and $D_2$.
 e) Order parameter of the $f$-wave phase. This order can also be interpreted
 as two $s$-wave paired hole states whose order parameters are out of phase
 by $\pi$.  f) Order parameter of the extended 
 $p$-wave phase, that appears in anisotropic lattices. 
}
\end{figure}

 In this section we consider a BFM on a triangular
 lattice. 
The geometry of the lattice under consideration here 
 is shown in Fig. \ref{phases}(a).
 This system is described by a similar
 Hubbard model as for the square lattice. 
 But now, besides  the triangular geometry,  
  we allow for
 two different values for the hopping amplitudes, for
   two types of lattice bonds, 
 as indicated in Fig. \ref{phases} a) by dashed
 and continuous lines.
$t_{f,a}$ and $t_{b,a}$ with $a=1,2$ are 
the fermionic and bosonic tunneling amplitudes between neighboring sites, 
where the index $a=1$ ($a=2$) refers to the continuous (dashed) bonds. 
 For the description of the isotropic case
 we equate $t_{b/f,1}$ and $t_{b/f,2}$, and
 define $t_f\equiv t_{f,1}=t_{f,2}$ and $t_b\equiv t_{b,1}=t_{b,2}$. 
$\mu_f$ ($\mu_b$) is the chemical potential for fermions (bosons), 
$U_{bb}$, $U_{ff}$, and $U_{bf}$ are the on-site boson-boson, fermion-fermion
and boson-fermion repulsion energy, respectively.

Just as for the case of a square lattice (Sec. VII), we consider the limit of weakly interacting bosons, in which
 the bosons form a BEC, for which we use  
  the same
 description. 
 The resulting  dispersion relation is now given by
 $\omega_\mathbf{k}  =  \sqrt{(\epsilon_{b,\mathbf{k}} -
 \epsilon_{b,0})(\epsilon_{b,\mathbf{k}} - \epsilon_{b,0} 
 +2U_{bb}n_b)}$,
%
 where the bare lattice dispersion is given by:
\bea\label{baredisp}
 \epsilon_{b,k} =  -t_{b,1} 2 \cos k_x - 
 t_{b,2}(2 \cos(k_x/2 + \sqrt{3}k_y/2)+ 2 \cos(k_x/2 - \sqrt{3}k_y/2)).
\eea 
For small values of $k_x$ and $k_y$, $\omega_{\mathbf{k}}$ can
 be expanded as:
 $\omega_{\mathbf{k}}  \sim  \sqrt{((2 t_{b,1}+t_{b,2})k_x^2 + 3 t_{b,2}k_y^2) 
 U_{bb}n_b}$,
 which gives us the two velocities
 $v_{b,x} = \sqrt{(2 t_{b,1} + t_{b,2})U_{bb}n_b}$
 and 
 $v_{b,y}  = \sqrt{3t_{b,2}U_{bb}n_b}$.

 
We again assume that these velocities of the 
condensate fluctuations are much larger than the Fermi velocity,  
     which corresponds to the conditions  
 $v_{b,x/y} > t_{f,1/2}$. 
 Therefore, large bosonic hopping amplitudes, 
 a bosonic density of $\approx 1$--$3$, and some intermediate
 value for $t_{f,1/2}$ will satisfy this requirement. 
 As before, the bosonic modes can be integrated out, and we obtain an approximately
non-retarded fermion-fermion interaction.  The induced potential $V_{{\rm
    ind.}, \mathbf{k}}$ is given by:
$V_{{\rm ind.}, \mathbf{k}}  =  - \tilde{V}/(1 + \xi_1^2 (2 - 2 \cos k_x) + 
 \xi_2^2(4 - 4 \cos(k_x/2)\cos(\sqrt{3}k_y/2)))$
with $\tilde{V}=U_{bf}^2/U_{bb}$, and $\xi_{a}$ are the healing lengths of
the Bose-Einstein condensate (BEC) 
and are given by  $\xi_{a}=\sqrt{t_{b,a}/2 n_b U_{bb}}$ with $a=1,2$. 
 We again arrive
 at a purely fermionic, non-retarded description of the 
 same form as before.
 This 
 is the effective model that we study with a numerical implementation 
 of the functional renormalization group.
\begin{figure}[h]
\centerline{\includegraphics[width=8cm]{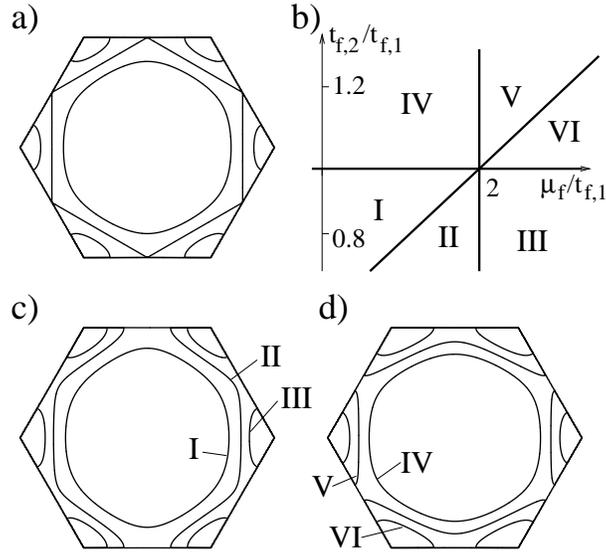}}
\caption{\label{FS}
a) Fermi surfaces for an isotropic lattice, for $\mu_f<2t_{f,1/2}$,
 $\mu_f= 2t_{f,1/2}$ (hexagonal shape), 
 and $\mu_f>2t_{f,1/2}$ (six disjoint arcs).
 b) Diagram of the different types of Fermi surfaces that can be created
 on an anisotropic lattice, by varying the ratio $t_{f,2}/t_{f,1}$ 
 and $\mu_f$.
 c) Fermi surfaces for $t_{f,2}<t_{f,1}$, for $\mu_f<4t_{f,2}-2t_{f,1}$, 
 $4t_{f,2}-2t_{f,1}<\mu_f<2t_{f,1}$, and $\mu_f>2t_{f,1}$, corresponding
 to the regimes I--III, respectively.
 d)
 Fermi surfaces for $t_{f,2}>t_{f,1}$, for $\mu_f<2t_{f,1}$, 
 $2t_{f,1}<\mu_f<4t_{f,2}-2t_{f,1}$, and $\mu_f>4t_{f,2}-2t_{f,1}$, corresponding
 to the regimes IV--VI, respectively.}
\end{figure}
 For the isotropic case, perfect nesting occurs at 3/4-filling, with three possible
nesting vectors: $\mathbf{Q}_1=(0,2\pi)$, $\mathbf{Q}_2=(\pi, \sqrt{3}\pi)$, and 
$\mathbf{Q}_3=(-\pi, \sqrt{3}\pi)$, leading to three different possible types
of instabilities per density wave channel.
For the anisotropic case, only $\mathbf{Q}_1$ can be a
 nesting vector, for the condition $\mu_f=2t_{f,1}$.
 To determine the 
 scale of the gaps, $\Delta$, associated with each of these 
order parameters, 
 we again use a
 'poor man's' scaling estimate,
 specifically: $\Delta \approx \Lambda_0 e^{-\ell_c}$, where $\ell_c$
is the point at which the RG flow diverges and the instability occurs. 

The RG is implemented numerically by discretizing the FS into $M$ patches. For the results shown in this Section, $M=24$ or $36$ was used. The CDW, AF and BCS channels are diagonalized at each RG step.
The dominant instability is the channel that has an eigenvalue (divided
 by the dimension of the matrix) with the
largest magnitude (for BCS one has to ensure that such eigenvalue is negative
so that the channel is attractive). Each element of the corresponding
eigenvector represents a given FS patch, and hence, the symmetry
of the dominant order parameter is reflected on the patch (i.e., angular)
dependence of each element around the FS. Using this method, we 
determine the phase diagram of the system in various limits.

We first consider  spin-$1/2$ fermions on an isotropic triangular lattice, i.e. 
with  $t_{f,1}=t_{f,2}\equiv t_f$. The FS for such a lattice behaves as
follows: For small filling the FS consists of one near-circular
piece, which then approaches the shape of a hexagon as $\mu_f$ approaches
the special value $\mu_f=2t_{f}$. At this special chemical potential, which
corresponds to $3/4$-filling, the FS is nested with the three
distinct nesting vectors ${\mathbf Q}_i$. For filling fractions larger than
$3/4$ the FS breaks into six disjoined arcs. 
 Examples for these different regimes are shown in Fig. \ref{FS} a).
Without coupling to
the BEC, the fermions form an $s$-wave pairing phase for attractive interactions, and
a Fermi liquid phase for repulsive interactions (ignoring high angular momentum 
pairing phases predicted by the Kohn-Luttinger theorem \cite{kohn} which would 
occur at energy scales much lower than the experimentally accessible
 regime), 
 except for the specific case 
$\mu_f = 2 t_{f}$,  where the system shows AF order for repulsive interactions.
A schematic picture of this order is shown in Fig. \ref{phases} b) 
for the nesting vector $\mathbf{Q}_1$. 
This behavior is similar to the one found for isotropic square lattice in the previous Section \cite{mathey06a}: $s$-wave pairing for attractive interaction,
 and Fermi liquid behavior for repulsive interaction, except 
 at an special filling, for which we find AF order due to nesting.
 An interesting difference for the triangular lattice
 is the three-fold degeneracy of the AF phase, an indication
 of frustration.

When the coupling to the
BEC is turned on, the isotropic triangular lattice shows a phase diagram of the type shown in Fig. \ref{PD}.
The $s$-wave pairing  phase slightly extends into the regime of positive $U_{ff}$,
because of the induced attractive interaction mediated by the bosonic fluctuations.
The regime that showed Fermi liquid behavior in the absence of the induced
interaction now shows a rich competition of various types of pairing.
In the regime where the density is below half-filling,
when the FS is approximately circular, the system shows $p$-wave pairing.
For fillings larger than $3/4$, when the FS consists of six disjoined parts,  
the fermions Cooper pair in a superconducting state with $f$ symmetry.
As shown in Fig. \ref{phases} e), the FS in this regime can also
be interpreted as two distinct near-circular Fermi surfaces of holes. 
In this interpretation each of these two fermionic systems is in an $s$-wave
pairing phase, but the relative phase between the two order parameters is $\pi$.
At $3/4$-filling and large values of $U_{ff}$, the system still shows AF
order. However, for smaller values of $U_{ff}$, and also for smaller
fillings, two phases with degenerate extended $d$ symmetry
develop. These superconducting orders have a sizeable $g$-wave component and
are approximately given by:
\begin{eqnarray}
\psi_{D_1} & = & \sin 2\theta + 0.5 \sin 4 \theta\\
\psi_{D_2} & = & \cos 2\theta - 0.5 \cos 4 \theta
\label{d1d2}
\end{eqnarray}
These order parameters are shown in Fig. \ref{phases} c) and d).
The shapes of the order parameters are energetically 
advantageous because, on the one hand, the order parameter maxima are located
at points at which the system has a high density of states 
(the 'corners' of the FS). Hence, when the superconducting gap opens, 
there is a large gain of condensation energy coming from these regions on
the FS.  On the other hand, the $d$-wave state
has lower kinetic energy than the $f$-wave, and hence is selected.

The phase diagram Fig. \ref{PD}
 has a number of 
similarities to the phase diagram for a BFM on a square lattice,
 such as the $s$- and the $p$-wave pairing phase, 
  and the existence of AF order 
 for a nested Fermi surface for large $U_{ff}$.
\begin{figure}[h]
\centerline{\includegraphics[width=8cm]{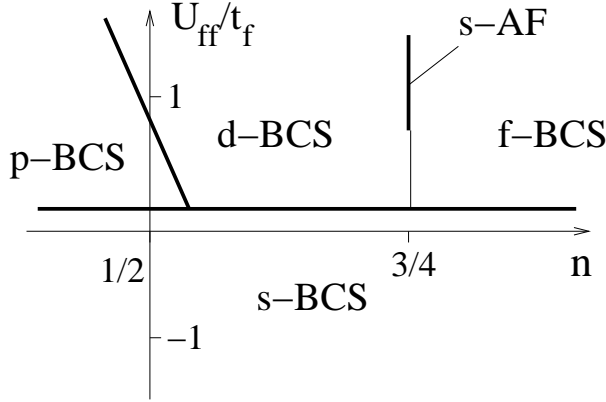}}
\caption{\label{PD}
Phase diagram
 of a Bose-Fermi mixture on a 2D isotropic triangular lattice.
 The vertical axis corresponds to the
  interaction strength, $U_{ff}/t_f$, whereas 
 the horizontal axis  corresponds to the
 filling fraction of the fermions per site, $n$. 
 The other parameters are given by $\tilde{V}/t_f = 3$, and $\xi_{a} = 1$
with $a=1,2$.}
\end{figure}
 However,
 the competition of pairing orders
 for positive $U_{ff}$ and
 intermediate  and large filling
 is much richer, due to the 
 more complex shape of the 
 Fermi surface. 

The energy gaps associated with these order parameters
 can be determined as we did in the previous Section \cite{mathey06a}, by using
 a 'poor man's' scaling argument.
 We find for the $s$-wave pairing and the AF order, that they are
 around $0.1 T_F$, 
where $T_F$ is the Fermi temperature of the system.
 For most of the exotic phases, we energy gaps of the order of
 $0.01 - 0.001 T_F$.

We now consider a BFM with spin-$1/2$ fermions on an anisotropic 
 triangular lattice, 
i.e. with unequal hopping amplitudes, $t_{f (b),1} \neq t_{f (b),2}$.
The shape of the FS behaves as follows:  For $t_{f,2}>t_{f,1}$, 
as one increases the chemical potential, the FS first breaks 
into four arcs at $\mu_{f}=2 t_{f,1}$, and then breaks into six arcs
at $\mu_{f} = 4t_{f,2}-2 t_{f,1}$,
 corresponding to the regimes IV--VI, in Fig. \ref{FS} b) and d). 
For $t_{f,2}<t_{f,1}$ the FS 
first breaks into two arcs at $\mu_{f} = 4t_{f,2}-2 t_{f,1}$, and then
breaks into six arcs at $\mu_f=2t_{f,1}$
 coresponding to the regimes I--III, in Fig. \ref{FS} b) and c). 
At the special chemical potential
$\mu_f=2t_{f,1}$ the FS is still nested,  but there is only one nesting vector 
along the direction of the bonds with hopping amplitude $t_{f,1}$.
In the absence of the coupling to the BEC the phase diagram has a similar
structure as for the isotropic case: $s$-wave pairing for attractive
interaction, Fermi liquid behavior for repulsive interaction, with the exception 
of the nested FS at $\mu_f=2t_{f,1}$ where one finds AF order (notice that
in this case the filling is not $3/4$).
  
\begin{figure}[h]
\centerline{\includegraphics[width=8cm]{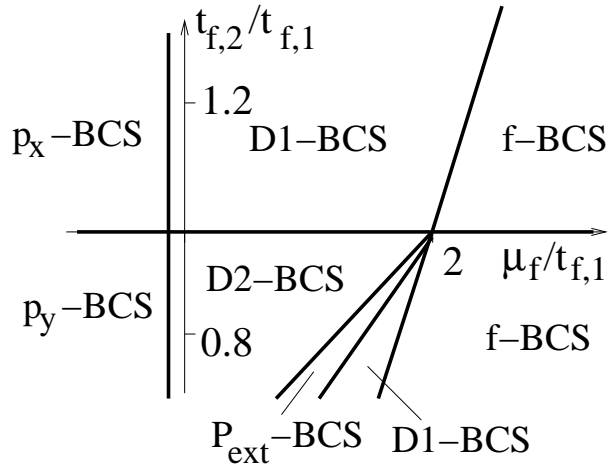}}
\caption{\label{PDaniso}
Phase diagram of  a Bose-Fermi mixture on an anisotropic triangular lattice.
 The vertical axis corresponds to
 the ratio $t_{f,2}/t_{f,1}$, the horizontal axis
 corresponds to the chemical potential $\mu_f$.  
 The other parameters are given by $\tilde{V}/t_f = 3$, 
 $U_{ff}/t_{f,1}=2$, and  $\xi_{a} = 1$ with $a=1,2$).}
\end{figure}

When the coupling to the bosons is turned on, one generates an even more
complicated competition of pairing phases for repulsive $U_{ff}$
in the vicinity of the point $\mu_f = 2t_{f,1}$, as is shown
in Fig. \ref{PDaniso}. Generally, for unequal hopping the degeneracy between 
$D_1$ and $D_2$ in (\ref{d1d2}), as well as $p_x$ and $p_y$ is lifted: In the
regime with $t_{f,2}>t_{f,1}$ ($t_{f,2}<t_{f,1}$), $D_1$ ($D_2$) and $p_x$
($p_y$) dominate. For $t_{f,2}>t_{f,1}$, in the intermediate regime, in which the
 FS consists of four disjoined arcs, 
 corresponding to the regime V in Fig. \ref{FS}, 
the type of ordering changes from $D_1$ to $f$.  
 For $t_{f,2}<t_{f,1}$, the type of pairing also eventually becomes $f$-wave,
 but first develops two other types of pairing, in
 the regime II in Fig. \ref{FS}. 
 Firstly, one finds an unusual
 extended $p$-wave symmetry, which is schematically shown in Fig. \ref{phases} f).
 Its wavefunction is of the form:
\bea
\psi_{P_{ext}} 
& = &\left\{
\begin{array}{cr}
 \sin^2 \theta  & -\pi/2 < \theta <\pi/2\\
 -\sin^2 \theta &  \pi/2 < \theta <3\pi/2
\end{array}
\right.  
\eea
The second type of pairing that appears before the system develops $f$-wave 
pairing is $D_1$. These unusual pairing states are energetically favorable
because of the 
 anisotropic shape of the FS. For the regime in which the FS has just 
barely broken up into two arcs, the order parameter assumes $p$-wave symmetry
and the maxima are located along the $y$-axis, where the density of states is highest.
As the region of open FS widens (see Fig. \ref{phases} f)), 
this pairing becomes energetically unfavorable, and
the system develops $D_1$-pairing, so that the maxima of the order parameter 
can again be located near the point of highest density of states. 
 The energy gaps associated 
 with these order parameters are
 of the same order of magnitude as  
 for 
 the isotropic lattice.

Finally, we consider a BFM with spinless fermions on an isotropic
lattice. Due to the absence of $s$-wave scattering between fermions
of the same spin state, there is no direct interaction, that is, $U_{ff}=0$.
Hence, in the absence of bosons, the spinless gas is non-interacting. 
The boson fluctuations, however, mediate an induced interaction between the fermions. 
Due to the anti-symmetry of the Cooper pair wavefunction, pairing occurs 
in an odd angular momentum channel.  We find a competition between $p$ and 
$f$-wave pairing symmetry. For small to intermediate filling ($n<0.65$), 
$p$-wave pairing dominates. For larger fillings, for which the FS first approaches
the shape of a hexagon and then breaks up into six arcs, the system
shows $f$-wave pairing. 
     Since  these larger fillings of fermions are typically realized in the center of an atomic trap,
 this result would suggest a comparatively easy way to create an exotic pairing state experimentally.  
In contrast to this,
 a spinless BFM on a square lattice
 only shows $p$-wave pairing, since for
 the quadrangular shape of its
 FS, channels of higher angular momentum 
 are of no advantage energetically. 


The many-body states discussed in this
 section can be observed through various
 methods: AF order could be revealed in 
time-of-flight images and Bragg scattering \cite{stenger99}, 
 noise correlations \cite{greiner05,foelling05,altman04,mathey05} can be used to detect 
the various pairing phases,
 laser stirring experiments \cite{raman99,onofrio99} can be used
 to detect the phase boundary between AF order and pairing. 
  The short-scale CDW fluctuations should give a signature 
 in a photo-association measurement. RF spectroscopy \cite{chin} can be used
 to quantify the gaps of the various phases.

\section*{\textcolor{NavyBlue}{VI. Conclusions}}
\addcontentsline{toc}{section}{VI. Conclusions}
In this article 
 we studied the phase diagrams of various
 low-dimensional ultracold atom systems.
 In Section III we studied atomic mixtures in one dimension,
 using Luttinger liquid theory. We argued
 that a Bose-Fermi mixture can be naturally looked at
 as a Luttinger liquid of polarons, and we discussed
 the rich phase diagrams of commensurate mixtures. 
 In Section IV, we studied the phases of two coupled two-dimensional superfluids
 at finite temperature. We found that that the critical temperature
 of the phase-locked phase is significantly increased 
 over its bare value, which we propose to use for realizing the
 Kibble-Zurek mechanism.
 When interactions between the two superfluids are present, we find
 additional phases which are partially superfluid, and partially disordered.
 In Section V, we used the powerful 
 method of functional renormalization group equations to determine the phases
 of Bose-Fermi mixtures in two-dimensional optical lattices. We found an 
 intricate competition of orders, including new types of exotic pairing for
  triangular lattices.
  In all these sections, 
 ideas how to probe our predictions were also given.



\section*{\textcolor{NavyBlue}{VII. Bibliography}}
\addcontentsline{toc}{section}{VII. Bibliography}

\noindent {\bf Primary Literature:}

\noindent {\bf Books and Reviews}
\bigskip

\noindent Leggett AJ (2001) Bose-Einstein condensation in the alkali gases: Some fundamental concepts. Rev Mod Phys 73:307

\noindent Bloch I, Dalibard J, Zwerger W (2007) Many-Body Physics with Ultracold
Gases. Rev. Mod. Phys. (to appear), arXiv:0704.3011v1

\vfill\eject

\begin{thebibliography}{99}
\bibitem{Definitions} 
Here we defined $K_{\beta}\equiv e^{2 \theta} \tilde{v}_f 
(\cos^2\psi/v_A + \sin^2\psi/v_a)$, $K_{\delta}\equiv  
K_b v_b (\sin^2\psi/v_A + \cos^2\psi/v_a)$,
$K_{\gamma}^{-1}\equiv  e^{- 2 \theta}/\tilde{v}_f 
(v_A \cos^2\psi + v_a \sin^2\psi)$,
$K_{\epsilon}^{-1}\equiv  K_b^{-1}/v_b (v_A \sin^2\psi + v_a \sin^2\psi)$,
${K}_{\beta\delta}= e^{\theta} \sqrt{K_b \tilde{v}_f v_b} 
\sin(2 \psi)/2 (1/v_a - 1/v_A)$, and
${K}_{\gamma\epsilon}^{-1}= e^{-\theta}/\sqrt{K_b \tilde{v}_f v_b} 
\sin(2 \psi)/2 (v_a - v_A)$. $\psi$ is given by
$\tan 2\psi = 4 \tilde{g} (v_b \tilde{v}_f)^{1/2}/(v_b^2 - \tilde{v}_f^2)$.
These expressions are obtained from diagonalizing $H_{\rm eff}$.
 Details will appear elsewhere.
\bibitem{endnote1} 
In this paper, we use $V_{f/b,\|(\perp)}$ to denote the optical lattice
potential experienced by fermionic/bosonic atoms in the longitudinal
(perpendicular) directions.
Independent tuning of the optical lattices for two
species of atoms can be achieved even with a single pair
of lasers by controlling the laser detuning and intensity separately.
\bibitem{endnoteCDW}
We note that in a homogeneous 1D system only quasi-long-range order
for CDW phase is possible. However, the inhomogeneous trap boundary
in a realistic experiment can pin the CDW phase to generate a true
density modulation.
\bibitem{kibbleremark}
 Since then, it has been established that cosmic strings can only 
 contribute a small fraction of the initial density perturbations. We
 thank T. Kibble for this comment.
\bibitem{trap}
Due to the confining trap the transition is, while
 visible, 'blurred' into a gradual cross-over, due to finite size effects, and,
 more importantly, due to the coexistence of several phases in the trap.
 This can also be expected for the phase transitions
 predicted in this paper.
\end{thebibliography}
\end{document}